\documentclass{aa}  

\usepackage{longtable}
\usepackage{graphicx}
\usepackage[colorlinks,pdfusetitle,urlcolor=blue,citecolor=blue,linkcolor=blue,bookmarksnumbered,plainpages=false]{hyperref}
\usepackage{adjustbox}

\usepackage{txfonts}
\usepackage{longtable,lscape}
\usepackage{booktabs} 
\usepackage{caption}
\usepackage{subfigure}
\usepackage{chngpage}

\def\R23{\mbox{$\rm R_{23}$}}

\begin{document}

\title{CLASH-VLT: Enhancement of (O/H) in z=0.35 RXJ\;2248-4431 cluster galaxies
\thanks{Based on observations obtained at the European Southern Observatory (ESO) Very Large  Telescope (VLT), Paranal, Chile; ESO large program 186.A-0798}}

\author{B. I. ~Ciocan\inst{1},
C.\,~Maier\inst{1},
B.\,L.~Ziegler\inst{1},
M.\,~Verdugo\inst{1}
}

\institute{University of Vienna, Department of Astrophysics, Tuerkenschanzstrasse 17, 1180 Vienna, Austria\\
\email{bianca-iulia.ciocan@univie.ac.at}
}

\titlerunning{MZR in cluster galaxies at $z \sim 0.35$}
\authorrunning{B. I. Ciocan et al.}

\date{Received ; accepted}

\abstract 
{}
{ Gas-phase metallicities offer insight into the chemical evolution of  galaxies, as they reflect the recycling of gas through star formation, galactic inflows and outflows. Environmental effects such as  star-formation quenching mechanisms play an important role in shaping the evolution of galaxies.  
Clusters of galaxies at $z<0.5$ are expected to be  the sites where environmental effects can be clearly observed with present-day  telescopes.}
{We explore the Frontier Fields cluster RXJ2248-443 at $z=0.348$ with VIMOS/VLT spectroscopy from CLASH-VLT, which covers a central region 
  corresponding to almost 2 virial radii. The fluxes of  $ [OII] \ \lambda 3727$, $H\beta$, $ [OIII] \ \lambda 5007$,  $H\alpha$  and $ [N II] \ \lambda 6584 $ emission lines were measured allowing the derivation of (O/H) gas metallicities, star formation rates based on extinction-corrected $H\alpha$ fluxes and  active galactic nuclei (AGN) contamination. We compare our sample of cluster galaxies to a population of field galaxies at similar redshifts.}
{We use the location of galaxies in projected phase-space to distinguish between cluster  and field galaxies.
Both populations  follow the star-forming-sequence in the diagnostic diagrams, which allow disentangling between the ionising sources in a galaxy, with only a low number of galaxies classified as Seyfert II.  Both field and cluster galaxies follow the "Main-Sequence" of star forming galaxies, with no substantial difference observed between the two populations. In the Mass - Metallicity (MZ) plane, both high mass field and cluster galaxies show comparable (O/H)s to the local SDSS MZ relation, with an offset of low mass galaxies ($log(M/{\rm M}_{\odot})<9.2$) towards higher metallicities. While both  the metallicities of "accreted" ($R<R_{500}$) and "infalling" ($R>R_{500}$) cluster members are comparable at all masses, the cluster galaxies from the intermediate, mass complete bin $(9.2<log(M/{\rm M}_{\odot})<10.2)$ show more enhanced metallicities than their field counterparts, by a factor of $0.065$ dex, with a $\sim 1.8 \sigma$ significance. The intermediate mass field galaxies are in accordance with the expected (O/H)s from the Fundamental Metallicity relation, while the cluster members deviate strongly from the model predictions, by a factor of $\sim 0.12$ dex. The results of this work are in accordance with studies of other clusters at $z<0.5$ and  favour the scenario in which the hot halo gas of low and intermediate mass cluster galaxies is removed due to ram pressure stripping, leading to an increase in their gas-phase metallicity. }{}


\maketitle



\setcounter{section}{0}
\section{Introduction}
Chemical abundances of galaxies represent an important tool to study galaxy evolution, as they reflect the complex interplay between star formation, gas outflows through winds and supernovae, and galactic gas inflows. A large number of studies concentrating on galaxies in the local universe have shown that there is a tight correlation of 0.1 dex between the stellar mass of a galaxy and its gas phase metallicity, giving rise to the so-called  Mass-Metallicity-Relation (MZR): the more massive the galaxy is, the higher its (O/H) oxygen abundance will be (e.g.  \citet{trem04}, \cite{kew2008}). There is also good evidence that this MZR evolves with redshift: for a given mass, gas-phase metallicities are higher at lower redshifts ( e.g. \cite{zahid13}, \cite{pm13}, \cite{maier15}).\\
Based on the extensive SDSS data, it has also been claimed that gas phase metallicities correlate with other galactic parameters as well, such as the star formation rate (SFR)  ( e.g. \cite{mannu10} ). \cite{salim14} has shown, that for local galaxies, the (O/H) anti-correlates with the SFR, regardless of the metallicity and SFR calibrators used.\\  
Observations have also demonstrated that  out to $z \sim 2$  there exists a “main sequence” (MS) of star forming galaxies, in which the SFR is closely linked with the stellar mass of a galaxy (e.g. \cite{brinchmann04},\cite{daddi07} ,\cite{whitaker14}). The MS has a characteristic specific SFR (sSFR) that declines weakly with the stellar mass as: $ sSFR \propto m_{star}^{\beta} \: with \: \beta \sim  -0.24$ \cite[]{renzpeng15}. Observations also demonstrate that the sSFR on the MS strongly evolves with redshift as:  $sSFR \propto (1 + z)^3 .$ For a given mass, the SFR has been decreasing at a steady rate by a factor of $\sim$ 20 from z = 2 to z = 0 (e.g. \cite{eblaz07}, \cite{peng10}).\\
An explanation to the origin of  the MZR and star-forming MS has been summarised
by  \citet[]{lilly13}, and is known as the bathtub model. The SFR in this gas-regulated model is closely linked to the mass of the gas reservoir in the galaxy, which is in term regulated by galactic gas inflows and outflows. The model  produces an implicit dependency of the gas-phase metallicity on both the stellar mass and the SFR, giving birth to the so-called Fundamental Metallicity Relation: $Z(M*,SFR)$, which is believed to be epoch-invariant  (e.g. \cite{mannu10}).\\
The evolution of galaxies is known to be internally regulated by the build up of the stellar mass and externally influenced by the environment in which they reside, e.g. \cite{peng12}.
Observations have shown that the physical properties  such  as the SFR, colour, morphology, metallicity, active galactic nuclei (AGN) fractions of  galaxies residing in a high density environment can differ from the properties of  isolated, field galaxies (e.g.  \cite{dressler80}, \cite{balgoh04}, \cite{kaufmann04}, \cite{ellison08}).  
A plausible mechanism which offers an explanation for the observed differences between the field and cluster populations (e.g. \cite[]{dressler80})  is the quenching of star formation when the galaxy is accreted into the cluster.  Cluster specific processes such as the hydrodynamical interaction between the hot intracluster medium (ICM) and the interstellar medium (ISM) of the galaxy or gravitative interactions between the cluster members can lead to a gradual cessation of SF.\\
Environmental effects are expected to get stronger at $z<0.5$ as the large scale structures of the universe evolve (e.g. Fig. 15. \cite{peng10}), and therefore, clusters of galaxies at intermediate redshifts should be the ideal sites to observe these effects, which effectively lead to the quenching of SF and which are, to some extent, expected to also affect the gas metallicities of cluster galaxies. E.g. both \cite{pasq12}  and \cite{pengmaio14} studied central and satellite SDSS galaxies and found that the average metallicity of satellites is higher than that of centrals, especially at low stellar masses. They speculated that low mass satellite galaxies,  which have a weaker gravitational potential than their massive counterparts,  are  more prone to environmental effects such as  the ram pressure stripping of the hot halo gas  (strangulation/starvation). \cite{cooper08} investigated a large sample of  SDSS galaxies and demonstrated that there exists a fairly tight relationship between
metallicity and environment, with  more
metal-rich galaxies found at regions of higher over-density.
\cite{maier16} and  \cite{maier19} studied the $Z(M*,SFR)$ of members of CLASH and LoCuSS clusters at $z<0.5$, in comparison to a sample of field galaxies, and observed more enhanced metallicities for low mass cluster members residing in over-dense regions than for their counterparts residing at larger cluster-centric radii and in the field. These enhanced metallicities were interpreted as the result of starvation. There is also indication for strangulation in $z>1$ cluster galaxies (\cite{maier19b}).
Strangulation/starvation is a mechanism that leads to the removal of the diffuse hot gas reservoir confined in the galaxy halo, while the cold gas-disdisck is left unperturbed.  This means that star formation can continue until the internal gas-reservoir is used up. Therefore, after the ram pressure of the ICM strips off  the hot halo gas, there will be a time delay until the system will undergo star formation quenching. As the inflow of pristine gas, which dilutes the ISM is truncated, the  metallicities of the retained gas will increase before the system undergoes complete quenching (see e.g. \cite{maier19} ). \\
In another recent publication, \cite{roberts19} used  X-ray observations for  $z < 0.1$ clusters to study the effects of the ICM on the quenching of satellite galaxies and found that   the fraction of quenched galaxies increases with the density of the ICM. Their results are consistent with the scenario in which cluster galaxies experience a slow quenching phase due to the removal of the hot gas confined into the galaxies halo followed by a rapid quenching mode associated with the removal of the cold gas component due to ram pressure stripping as the galaxies approach the cluster centre.\\
 In this paper we further explore the environmental effects on gas regulation within galaxies by measuring (O/H) abundances in a very massive cluster - RXJ2248-443 - at $z \sim 0.35$, and by comparing them to chemical abundances of field galaxies at similar redshifts. We investigate two connected relations: the Mass-Metallicity (MZR) and Mass-sSFR relation  to establish whether they have an environmental dependency. The chemical abundances were derived from several 
diagnostics based on strong emission lines (ELs) such as:  H$\beta$, ${\rm [O\,III]\,}{\lambda\,5007}$, H$\alpha$, and [NII]${\lambda\,6584}$.
At higher redshifts (z $\sim 0.5$), these ELs are shifted out of the wavelength range of optical spectroscopic surveys, with H$\alpha$ and [NII]${\lambda\,6584}$ falling at near-infrared wavelengths, requiring thus near-infrared spectroscopy to observe them.  Therefore, the epoch we probe ($z<0.5$) is the highest redshift at which all these ELs can be observed 
with the same optical spectrograph.\\
This paper is structured as follows: in Sect. 2 we present the selection of the cluster CLASH-VLT EL galaxies at $z \sim 0.35$ and their VIMOS spectroscopy. We also introduce  the comparison samples of field galaxies from CLASH-VLT and  zCOSMOS. We investigate the  stellar masses, the active galactic nucleus (AGN) contribution and present the
derivation of SFRs and gas-phase metallicities for the 121 cluster and 130  field  galaxies that we observe. In Sect. 3 we present the mass - sSFR  relation and the mass-metallicity relation for cluster and field galaxies at $z\sim0.35$. We investigate how the dense cluster environment affects the chemical enrichment of the galaxies. In Sect. 4 we discuss whether the mechanism of ram pressure stripping of the hot halo gas can account for the observed metallicity trends implied by our results and we present a comparison with literature metallicity studies of the intermediate redshfift universe. In section 5 we summarise our conclusions. \\
Throughout this study,  we have used the concordance  $\Lambda CDM$ cosmology with $ H_{0} = 70\: km \:s^{-1}\: Mpc^{-1}$, $\Omega_{0}  = 0.25$, $\Omega_{\Lambda} = 0.75$. We assume a Salpeter \cite{salp55} initial mass function (IMF) for all the derived stellar masses and SFRs. We also note that (gas phase) metallicity and abundance are used to denote oxygen abundance, (O/H), throughout this paper.

\section{Data and measurements: cluster and field galaxies at $z\sim 0.35$ }
\subsection{Cluster and field galaxies from CLASH-VLT}
The CLASH RXJ2248-4431 cluster (hereafter R2248) also known as Abell S1063, is a massive
X-ray-luminous galaxy cluster at $z \sim 0.35$. 
The system was  observed as part of the HST Frontier Fields project and also as part of the CLASH ("Cluster Lensing And Supernova survey with
Hubble", \cite{postman12}) and CLASH-VLT survey (\cite{rosati14}). Chandra X-ray observations of R2248 revealed a very high X-ray luminosity as well as a  very hot intracluster gas with a temperature of up to $\sim14.3$ keV (\cite{gomez12}). Observations have also shown that the cluster members have high velocity dispersions in the range of over 1000 $ km s^{-1}$. Both the
high velocity dispersion and  X-ray temperature suggest that Abel S1063 is a very massive cluster with an estimated mass, according to \cite{gomez12}, of  $ M_{200} > 2.5 * 10^{15 }{\rm M}_{\odot}$  that probably formed through a merger event.  \\
The CLASH-VLT survey is the follow-up spectroscopic campaign of CLASH on 13 intermediate redshift $ 0.3 < z < 0.6$ clusters accessible from the Very Large Telescope (VLT), at Cerro Paranal in the Atacama Desert of Chile.  The survey had 225 hours allocated for observations, with 200 hours of multi-object spectroscopy and 25 hours of pre-imaging, primarily using the VIMOS spectrographs low-resolution LR-blue grism and the middle-resolution MR-grism.
The data set on which this work is based consists of 947 CLASH-VLT VIMOS spectra, for both field galaxies around R2248 and  cluster members. The galaxies span a redshift between $0.01<z<0.9$. The observations  were obtained as part ESO large program 186.A-0798 (\cite{rosati14}, Rosati et al. in prep.).  The spectra were registered with the VIMOS spectrograph with an exposure time of 1 h,  using 4 middle-resolution masks  with 1 arcsec slits and a  resolution  of R=580.  The covered  wavelength range is  4800-10000$\AA$,
thereby including ELs from $ [O II]\: \lambda\: 3727$ to $ [N II]\: \lambda\: 6584 $ for 
$z<0.5$ galaxies, and $ [O II]\: \lambda \:3727$ to  $[OIII]\: \lambda \:5007$ for  $0.5<z<0.9$.  The observations were carried out at a wavelength domain contaminated by only a few strong OH sky line. The VIMOS slit-masks were
designed into sets of four pointings, with one of the quadrants centred on the cluster core.
 The data reduction of   947 galaxy spectra, as well as the redshift determination, was done with VIPGI, the VIMOS Interactive Pipeline and Graphical Interface (VIPGI; \cite{scod05}) software package, by the CLASH-VLT team.\\ 
Additional observations for the  CLASH-VLT target R2248 are archival imaging data provided by  the ESO Wide Field Imager (WFI) of the 2.2-m MPG telescope of La Silla Observatory. We have used a catalogue containing the observed UBVRIz magnitudes  together with the errors for each filter for 636  field and cluster galaxies. The photometric data was corrected for galactic extinction according to \cite{schlaflyf11}.\\
In addition to the CLASH-VLT galaxies, we have used a sample of zCOSMOS field galaxies with $0.373 < z < 0.413$ as a comparison to R2248, due to the low number of field galaxies with 0.3<z<0.4 from our data set. This zCOSMOS sample corresponds to the one used as a comparison to the MACSJ0416 cluster in \cite{maier16}.\\
For our metallicity study, we have also used a local comparison sample of SDSS emission line galaxies in the redshift range 0.04 < z < 0.08, as in  \cite{maier16}.   AGNs and LINERs were excluded from the sample based on the BPT (\cite{bald81}) diagnostic. We have applied a S/N  threshold in $H\alpha$ >10 for this sample of local galaxies  and computed the oxygen abundances using the O3N2 method of \cite{kewley13a}. Therefore, we point out that this sample is different from other local comparison SDSS samples often used in the literature, such as the sample of,  e.g., \cite{trem04}.   

\subsection{EL fluxes}
The EL fluxes of R2248 cluster members and field galaxies were measured interactively using the package splot in IRAF,  adopting the same approach as described in \cite{maier16}.  We only measure the EL fluxes for the  middle-resolution spectra, because in the low-resolution spectra only the [OII] $\lambda3727$ EL is present, making the derivation of oxygen abundances impossible.
The ELs of interest: $ [O II]$, $H\beta$, $ [O III] $,  $H\alpha$  and $ [N II] $ were each individually fitted with a single Gaussian profile. This was done by marking two positions, blueward and redward of the EL, which give the region to be fitted as well as the fixed linear continuum.  Through this fitting procedure, we obtain a value for the centre, the continuum at the centre, the integrated flux and the equivalent width. Except for the continuum, all the aforementioned parameters are based on the fitted analytic profiles. This procedure was applied 3 times for each EL of interest, in order to derive the mean flux value and  the errors of the flux measurement. The flux errors were dominated by systematic uncertainties in establishing the local continuum, which was conservatively estimated by exploring rather extreme possibilities. \\ When compared to theoretical values, the observed ratios of $\frac{H\alpha}{H\beta}$ constrain the degree to which the stellar atmospheric absorption lines reduce the measured Balmer emission, as well as  the amount of reddening induced by dust. Thereby, spectra with high signal to noise are required. In the absence of very high-quality Balmer line measurements, statistical corrections have to be applied to the measured strength of the Balmer emission lines, particularly the $H\beta$ emission line. Following the
recommendation of \cite{kobul99}, we assume an average underlying stellar absorption in $H\beta$ of 3 $\pm$ 2 $\AA$ and correct the equivalent width of the spectral line by this amount, thereby increasing the $H\beta$ line flux. This correction is needed mainly when the EL is weak compared to the continuum. Such corrections of the  $H\beta$ emission line flux are often used in literature for data at similar redshifts and with similar spectral resolution as the one used within this work (e.g. \cite{maier05}, \cite{maier16}).
Quality flags were assigned to all the $947$ investigated galaxy spectra. If a spectrum is described by a quality flag \textit{f=1}, then all the EL fluxes could be measured. If a spectrum is described by a quality flag  \textit{f=2}, then the measurement of one EL flux was problematic: either the EL was very weak or the respective line was an artefact due to  sky-lines. If a  quality flag \textit{f=3} is assigned to a spectrum,  then two or more EL fluxes could not be measured. Most galaxies described by a quality flag \textit{f=3} are earlier-types showing few/no ELs.\\ A consistency check on the EL flux measurements was done by comparing the IRAF measurements to the measurements offered by FADO ("Fitting Analysis using Differential Evolution Optimisation" - \cite{papaderos17}), which is a tool specially designed to perform population spectral synthesis  to
derive different physical parameters from a galaxy spectrum, including EL fluxes. It is worth mentioning that the fluxes offered by FADO are corrected for underlying stellar absorption.  We observe a good agreement between the measurements offered by the two different tools, however, we have chosen to use the IRAF flux measurements as FADO had problems fitting the template SEDs for more than half of the provided spectra, due to the high quality requirements  which were not met with middle-resolution observations. The flux measurements are given in Table \ref{tab}.

\subsection{Selection of the cluster and field galaxies for our metallicity
study - Phase Space diagram}
\begin{figure*}
\begin{subfigure}
    \centering
    \includegraphics[width=0.7\textwidth,angle=0,clip=true]{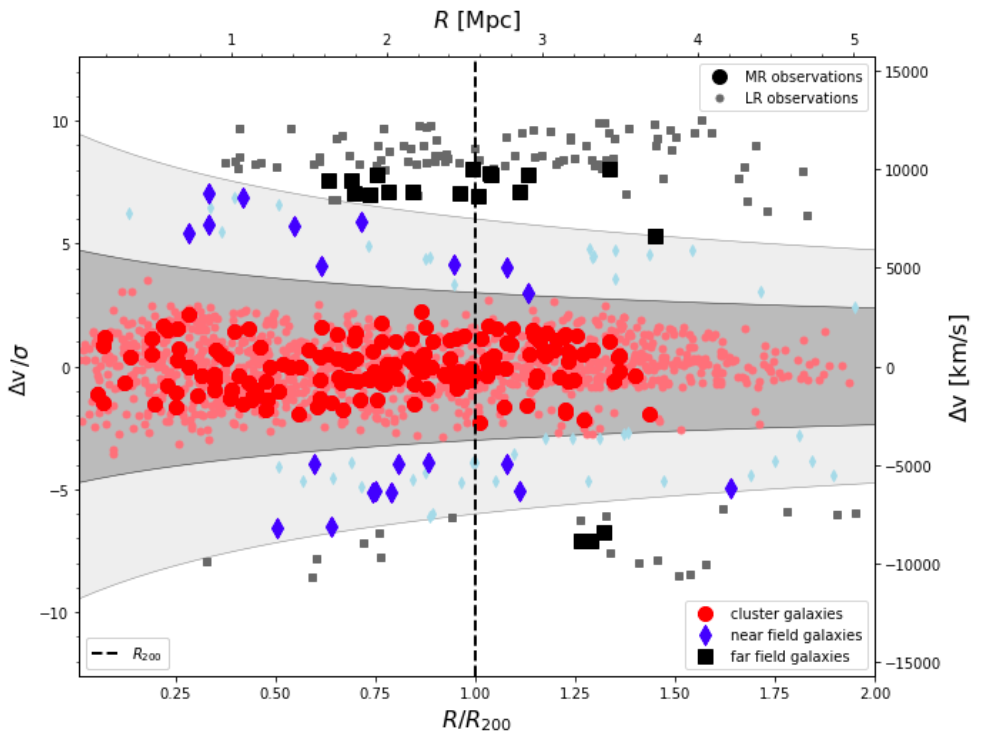}
    \centering
\end{subfigure} 
\begin{subfigure}
    \centering
    \includegraphics[width=0.35\textwidth,angle=0,clip=true]{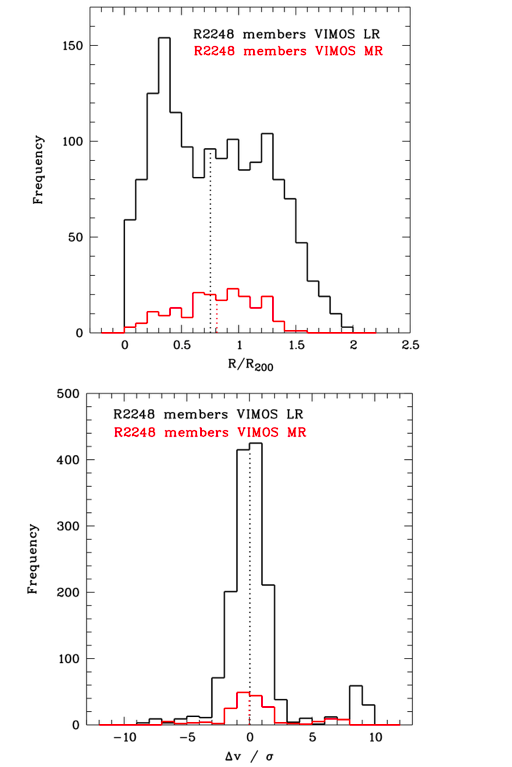}
    \centering
\end{subfigure} 
   \caption{\textit{Left}: Cluster centric radius vs. line-of-sight velocity for the sample of CLASH galaxies with $0.3<z<0.4$. Both galaxies with middle resolution and low resolution observations are shown. The red points which fall within the $3\sigma$ contours of the trumpet shaped caustic  profile (dark-grey area) are classified as R2248 cluster galaxies. The blue diamonds, located between  $3\sigma$-$6\sigma$ (light-grey area) contours are considered to be galaxies from the near field, whereas the black squares, which are located beyond the $6\sigma$ contours of the caustic are classified as far-field galaxies. Only the red symbols are considered for the computation of the cluster specific parameters such as $R_{200}$ and $M_{200}$. The dashed black line shows $R_{200}$. The large symbols represent the galaxies with middle resolution observations while the small ones represent the galaxies observed with a low resolution grism, for which no flux measurements  of other ELs than [OII] are available.\\ \textit{Right-top}: Histogram showing the cluster centric radius normalised to $R_{200}$ for the R2248 cluster members with low resolution observations (black) and middle resolution observations (red). The dotted lines show the median value for $\frac{R}{R_{200}}$ of both samples, colour coded accordingly.  \\ textit{Right-bottom:} Histogram showing $\Delta$v, normalised to the clusters velocity dispersion $\sigma$ for the R2248 cluster members with low resolution observations (black) and middle resolution observations (red). The dotted lines represent the median values for $\frac{\Delta v}{\sigma}$ of both samples, colour coded accordingly.  } 
\label{PhaseSpacewhole}
\end{figure*}

Position vs. velocity diagrams  offer great insight into the orbital histories of the cluster members. The location of galaxies in such projected phase space diagrams, i.e. cluster centric radius vs. line-of-sight velocity, provides information about the accretion history of the systems. Virialised galaxies, that have been in the cluster for  a few crossings, will accumulate  at low cluster centric radii and will have on average low velocities. The infalling or recently accreted cluster galaxies will be found at higher velocities overall, and will be thus spatially separated from the accreted members. Different recent studies have used such phase-space diagrams  in order to probe both the kinematics and the accretion epochs of the cluster  members, as well as  to constrain the density and mass profile of the cluster (e.g. \cite{jaffe15}, \cite{capasso19}, \cite{haines15}).\\
The R2248 members were identified from the cluster centric radius vs. line-of-sight velocity plane as lying within the “trumpet”-shaped caustic profile expected for galaxies infalling into the system
and subsequently orbiting within the massive virialised structure, by adopting the mass profile of model of \cite{carlberg97} . This method is based on a theoretical model which assumes that clusters are singular isothermal spheres. The mass model works as follows:  first, the difference in velocity $\Delta$v between the mean velocity of the cluster $c \bar z$ and the velocity of each individual galaxy is computed. Afterwards, the  values of $\Delta$v are normalised to the clusters velocity dispersion $\sigma$ and  plotted against the projected cluster centric radius of the individual members in units of $R_{200}$. The parameter  $R_{200}$ represents  the
radius where the density is 200 times the mean density of the universe, at the clusters' redshift. The velocity dispersion profile used to mark the caustics within this model is calculated as follows:
$
\sigma^2=B\cdot \frac{c_{1} r /(1+c_{1}r)+c_{2}}{1+r/b} \label{sigmar}
$
where $B=1/4$ and $b=0.66$ are two parameters adjusted to fit the observed projected velocity dispersion at R. The $c_{1}$ and $c_{2}$ parameters are fixed to be 0 and 1, respectively. This model is used to mark the $3\sigma$ and $6\sigma$ contours in the $\Delta$v  - $R_{200}$ plane, which separate between cluster members -located within the $3\sigma$ contours of the caustic profile- and field galaxies -located beyond $3\sigma$ (see more details in \cite{carlberg97}).\\
This theoretical model was implemented in a python code, and for the computation of the cluster specific parameters we have used the entire R2248 catalogue, which  includes both the low resolution and middle resolution spectra. We have only used the galaxies from the R2248 catalogue, which have  reliable redshift estimates, based on the secure identification of at least two spectral features, i.e. absorption and emission lines.\\ 
Fig. \ref{PhaseSpacewhole} shows on the left hand side the phase-space diagram for all galaxies from the R2248 catalogue with redshifts between $0.3<z<0.4$.  The red points located within the $3\sigma$ contours of the caustic profile (dark-grey area) are classified as  being members of the R2248 cluster. The blue diamonds, located between  $3\sigma$-$6\sigma$ (light-grey area) are classified as galaxies from the near field, while the black squares, which are located beyond  $6\sigma$ (white area)  are considered to be far-field galaxies. The small, faintly coloured  symbols show the galaxies with low resolution observations while the bigger, brighter ones show the galaxies with middle resolution observations, i.e. the data sample on which this work is based.  The low resolution spectra are not suited for a metallicity study,  as  only the [OII] $\lambda3727$ EL is present in these spectra, making the derivation of oxygen abundances impossible. The dashed, black line in the phase space diagram represents $R_{200}$.\\ The top panel from the right-hand side  shows the distribution of the R2248 galaxies with middle resolution  (red) and low resolution (black) spectra according to their cluster centric radius, normalised to $R_{200}$. The dotted lines give the median values of the cluster centric distances of the 2 samples, colour coded accordingly. Our sub-sample of middle resolution R2248 galaxies shows a similar distribution to the parent low resolution sample, but the middle resolution observations only reach a maximum cluster centric radius of $R=1.5 \cdot R_{200}$ in contrast to the low resolution observations which cover larger cluster-centric radii, of $R=2 \cdot R_{200}$. The distribution of cluster members with low resolution observations however peaks at a cluster centric distance of $\sim 0.3 \cdot R_{200}$, whereas the distribution of cluster members identified with the middle resolution observations peaks at a cluster centric distance of nearly  $R_{200}$.  This can be explained by the fact that the low resolution observations were designed as such, to observe mostly the crowded, central region of the cluster, in order  to identify passive members as well as gravitational lensed galaxies. Due to the high multiplexing capacity of the VIMOS low-resolution mask,  these observations allowed more redshifts estimates in the central region of the cluster, compared to the middle-resolution mask. The middle resolution follow up observations were designed in only a few pointings spread around the field of R2248. \\
The bottom panel from the right hand side shows a histogram with the distribution of $\Delta$v, normalised to the clusters velocity dispersion $\sigma$ for the R2248 cluster members with low resolution observations (black) and middle resolution observations (red). The median values for $\frac{\Delta v}{\sigma}$ for the two samples are given as the dottes lines, colour coded accordingly. The sub-sample of middle resolution observations shows a fairly similar distribution to the parent sample of low-resolution spectra in the $\frac{\Delta v}{\sigma}$  space.\\According to this mass model, the R2248 cluster can be described by  the following parameters:\: $z_{cluster}=0.3457 \pm 0.0001$\:, cluster members $\sim 1361,$
$R_{200}=2.57 \pm0.05$ Mpc\:,
$R_{500}=1.71 \pm 0.05$ Mpc\:,
$M_{200}=1.77 \pm 0.03 \cdot 10^{15} {\rm M}_{\odot} $\:,
$\langle v \rangle = 1180.916 \pm 40$ km/s.
The errors for the cluster parameters were computed through the so-called Jackknife resampling method by subsequently removing a random sub-sample of $\sim 200$ galaxies from the R2248 catalogue containing both the low resolution and middle resolution observations and running the code several tens of times. The mean values and the errors for $R_{200}$, $M_{200}$ and $\langle v \rangle$ were then computed from the values offered by each run of the code with the reduced version of the R2248 catalogue. One should, however, keep in mind that these are just the random errors and do not represent the true uncertainties for the derived values of the cluster specific parameters. \\ 
These values are in fairly good agreement (within the errors) with other studies of R2248.  \cite{gomez12} computed the mass of R2248 from the $M_{200} -\sigma_{DM}$ scaling relation and obtained the following values: $ M_{200} = 3.97_{ -0.9}^{+1.6} \cdot 10^{15} {\rm M}_{\odot}$
and an $R_{200} = 3.15+0.3 Mpc$, but based on only 51 cluster members, for which they could measure the velocity dispersion.  \cite{melc15} study the weak lensing masses and the galaxy distributions in massive clusters based on data from the Dark Energy Survey, and they derived a radius for R2248 of  $R_{200}\sim 2.2 Mpc$. On the other hand, in a  publication by \cite{pizz17}, the authors perform
a maximum likelihood kinematic analysis on the RXJ2248 cluster to determine the total mass profile in modified gravity models, assuming a spherical Navarro-Frenk-White profile, and obtained a value for $R_{200} = 2.7 \pm 0.12 Mpc$.  It is clear that different methods yield slightly different values for the cluster-specific parameters. \\
The investigated cluster galaxies with middle resolution observations seem to be located, in projection,  close to the cluster core.  This is however not surprising,  as the R2248 cluster is known to be one of the most massive CLASH clusters. The observations for the CLASH-VLT clusters were carried out by using 8-12 VIMOS pointings, with one quadrant locked onto the core, and the data set on which this work is based corresponds to the observations from  the crowded central region of the cluster. Therefore, most galaxies don not reach the limits of the caustic profile. \\
The R2248 cluster galaxies span a redshift $0.33<z<0.36$, and from the parent middle resolution sample of 947 galaxies, 178  have a redshift in the aforementioned range, out of which only 61 have flux measurements for all ELs of interest (described by a quality flag \textit{f=1}). In order to have enough objects for the statistics, both  \textit{f=1} and  \textit{f=2} cluster galaxies are used throughout this study, with our final sample consisting  of 121 cluster members. \\
The comparison sample of CLASH field galaxies was restricted to a redshift range $0.3<z<0.4$ in order to avoid any biases with galaxy evolution, and only the \textit{f=1} field galaxies were considered. As only 37 CLASH field galaxies are in the aforementioned redshift regime, having flux measurements for all ELs of interest, an additional sample of  93 zCOSMOS field galaxies with $0.373 < z < 0.413$ from \cite{maier16} was used as a comparison to the R2248 cluster galaxies. \\To select a sample of galaxies for the metallicity study, some constraints on the signal to noise (S/N) were applied. The S/N in the $H\alpha$ line was set to be $S/N_{H\alpha}>5$. For the [OII], $H\beta$, [OIII] and [NII] emission lines, the S/N threshold was restricted  to be $S/N>2$. The rest-frame  equivalent widths of $H\beta$ and $H\alpha$ were chosen as such: EW$(H\beta) > 2\AA$ and  EW$(H\alpha) > 3\AA$. This results in  90$\%$ of our cluster galaxies to have  $S/N_{H\alpha}>10$, and 10$\%$ with $5<S/N_{H\alpha}<10$.\\

\subsection{Stellar masses}
\label{sec:masses}
Stellar masses  were derived using
the code  LePhare of \cite{arnilb11}, which fits stellar population synthesis models to the
available photomety: the observed magnitudes in the WFI UBVRIz-bands. The code is composed of a set of FORTAN commands, which incorporate the standard $\chi^2$ minimisation method, that offers the best match to a reference set of spectral templates for the given photometric data. The used population synthesis models are based on the galaxy library of SSP spectra from \cite{bruzcharl03}, and the templates were fitted for stellar ages between 1 Gyr and 11 Gyrs. This premise was based on the fact that at $z\sim 0.35$ , the universe was $ \sim 9.8$ Gyrs old and most of the stellar mass was assembled by that time. The redshift  was kept fixed and  the number of extinction E(B-V) values was limited to: 0.1, 0.3, 0.5 in order to avoid over-fitting. The used cosmological parameters are in accordance to the $\Lambda CDM$ model, and based on them, the code will reject models older than the age of the universe. The absolute magnitudes were derived in the reference band, chosen to be the R-band, from the apparent magnitude in the  observed band, which was selected automatically to be as close as
possible to the reference band, redshifted in the observer frame, in order to limit the template dependency. LePhare assumes a \cite{chabrier03}  Initial Mass Function (IMF) when deriving the stellar masses.
These Chabrier masses were then converted to Salpeter  IMF (\cite{salp55}) masses according to
\cite{pozzetti07}, who found the factor of 1.7 to be a systematic median offset  in the masses derived with the two different IMFs. This factor was also found to have a very small dispersion and a rather constant value for a wide range of SFHs.
We are confident that the derived stellar masses are robust, since
the B - R colour encompasses the redshifted $4000\AA$ break and is thus sensitive to the galaxy mass-to-light ratios (e.g. \cite{kaufm03b}).
The values for the stellar masses of the R2248 cluster galaxies are given in Table \ref{tab}.

\subsection{Star formation or AGNs?}
\label{sec:BPT}
Using a set of 4 strong emission lines : [OIII], [NII], $H\alpha$, and $H\beta$, one can reliably distinguish between star-forming galaxies, Seyfert galaxies, low ionisation nuclear emission regions (LINERs) and composite galaxies with both star forming regions and an active galactic nucleus. This diagnostic diagram, which allows to asses whether the main source of ionisation is of stellar origin or rather associated with AGN activity,  was first introduced by \cite{bald81}  (BPT diagrams).  The panel from the left-hand side of Fig. \ref{BPTfc} shows a BPT diagram for the R2248 cluster galaxies while the panel from the right-hand side shows the same, but  for the comparison sample of CLASH field galaxies with $0.3<z<0.4$. The colour coding corresponds to the oxygen abundance, as derived using the O3N2 method of \cite{kewley13a}. The blue solid curve in both diagrams represents the theoretical curve of \cite{kewley01}  and the red dashed one the empirical curve of \cite{kaufm03}, which separate SF galaxies from AGNs. The green solid curve of  \cite{sw17}  depicts  the separation curve between Seyfert II galaxies and LINERs.  If galaxies lie above and to the left of the 3 curves, with high values in both $[NII]/H\alpha$ and  $ [OIII]/H\beta$, then they can be classified as Seyferts. If the  galaxies lie above the two curves of \cite{kewley01} and \cite{kaufm03}, and below the curve of \cite{sw17},  having high values only in $[NII]/H\alpha$, then they are classified as LINERS. For galaxies that lie below and to the right of the 2 borderline curves of \cite{kewley01} and \cite{kaufm03}, the main source of ionisation comes from the stellar component, making them star forming galaxies.  The area between these 2 curves represents the "uncertainty" region populated by composite galaxies with both active star formation and an AGN.\\ Most galaxies from our sample follow the SF sequence in the BPT diagnostic diagram, and all galaxies which are classified as Seyfert II or LINERS were excluded from the metallicity study (a total number of 10 cluster galaxies and 1 CLASH field galaxy). According to \cite{maier16}, Fig. 2,  all the zCOSMOS field galaxies can be classified as SF galaxies, and thus, none were excluded from the metallicity study. \\

\subsection{SFRs}
\label{sec:SFRs}
The SFRs for both the cluster members and  the comparison sample of field galaxies were  computed from the extinction corrected luminosity of the $H\alpha$ line.\\
The $H\alpha$ EL is one of the best-understood and  most reliable SFR indicators, as this nebular emission arises directly from the recombination of HII gas ionised by the most massive O- and early B-type stars and, therefore, traces the star formation over the lifetimes of these stars. Only the most massive stars with masses $>10\:{\rm M}_{\odot}$  and lifetimes $< 20\: Myrs$ have a significant contribution to the ionising flux, so the ELs, which re-emit this luminosity,
provide  us with an instantaneous estimate of the SFR, independent of the previous star formation history. The luminosity of the $H\alpha$ emission line can be used to calculate the SFR, by applying the \cite{ken98} conversion: 
\begin{equation} SFR({\rm M}_{\odot} \cdot yr^{-1})=7.9 \cdot 10^{-42} L(H\alpha)(ergs/s)  .\label{eq:5.3.1}\end{equation}
The intensities of ELs arising from gas nebulae are strongly affected by selectively absorbing material on the line of sight to the observer. Thus, the most important source of systematic error in $H\alpha$-derived SFRs is the extinction induced by dust. In order to accurately estimate SFRs, the luminosity of the $H\alpha$ EL was  corrected
for extinction based on the Balmer decrement, following the equations introduced by \cite{brock71}.\\ In order to account for the slit-losses, as the VIMOS masks have slits of 1 arcsec,  an aperture correction was applied to the derived SFRs. Each VIMOS spectrum was convolved with the WFI R-band filter and then, this magnitude was compared to observed R-band magnitude of the respective galaxy. The
difference between the two magnitudes offers the aperture correction factor for each spectrum. For $\sim 90 \%$ of the sample, the aperture correction factors have values between 1 and 3. This correction assumes that both the $H\beta$ line flux and R-band continuum were equally affected by the slit losses and that the EW of $H\alpha$ remains constant throughout the entire galaxy (see e.g. \cite{maier09}).\\ The main advantages of this calibration are its high sensitivity, and the direct proportionality between the luminosity of the nebular emission and the SFR. \\
The values for the sSFRs of the R2248 cluster galaxies are given in  in Table \ref{tab}.
\\  

\subsection{Oxygen abundances}
\label{Oxabund}

The oxygen abundances for the R2248 cluster members as well as for the comparison sample of field galaxies were computed by means of 3 empirically calibrated metallicity estimators based on the relative strengths of ELs: the calibration by \cite{kewley13a} which uses the O3N2 index  and the calibrations derived by \cite{maier16} for SF galaxies using the $[OIII]/H\beta$ and the $[NII]/H\alpha$  EL ratios.\\
The O3N2 index was first introduced by \cite{alloin79}  as: \begin{equation} O3N2=log([O III] \lambda5007/H\beta)/([N II] \lambda6583/H\alpha) \end{equation}
 By means of stellar evolutionary synthesis and photoionisation models with chemical evolution measurements from cosmological hydrodynamic simulations,  \cite{kewley13a} derived the following empirical metallicity calibration based on the O3N2 index: \begin{equation} 12 + log(O/H)= 8.97 - 0.32 \cdot O3N2 \end{equation} This calibration corresponds to  the \cite{petpag04} O3N2 calibration for the \cite{kewdop02} metallicity scale, with a difference in the normalisation factor of $+0.24$. As this calibration relies on ratios of ELs which are close in wavelength, no corrections for reddening, nor accurate flux calibrations of the spectra are needed. \\ 
\cite{maier16} derived two metallicity calibrations for galaxies on the star-forming  sequence  based on equations 3. and 4. of  \cite{kewley13a}. The calibration as a function of the line ratio $[OIII]/H\beta$ is very useful as a metallicity indicator for high-redshifted galaxies which do not have measurements of the $[N II]$ and $H\alpha$ ELs, and it can be expressed as follows: \begin{equation} 12+ log(O/H) = 8.97 - 0.32 \cdot
(log([OIII]/H\beta) - 0.61/\end{equation}$$(log([OIII]/H\beta) - 1.1) + 0.08) $$
The (O/H) calibration as a function of the  $[N II]/H\alpha$ line ratio is given by the following equation: \begin{equation}
12+ log(O/H) = 8.97 - 0.32 \cdot (1.1 + 0.61/(log([NII]/H\alpha) +\end{equation} $$+ 0.08)- log([NII]/H\alpha)) $$This equation is similar to the (O/H) calibration of \cite{salim15} , which uses the re-calibrated N2 index, that matches the \cite{petpag04} O3N2 metallicity calibration. It was however demonstrated that the N2 metallicity estimator saturates for high (O/H) values.\\ The derived (O/H)s  for both cluster and field galaxies using the 3 different calibrations are in pretty  good agreement to each other. Recent  studies of blue (\cite{bresloin16}) and red supergiants (\cite{davies17}) have, however, demonstrated  that the metallicity calibration based on strong emission line ratios, yielding the most accurate, absolute metallicities is the O3N2 calibration, and we have chosen to use this calibration throughout this study, when all 4 ELs of interest were present. 
The values for the (O/H)s of the R2248 cluster galaxies are listed in  in Table \ref{tab}.

\section{Results:Mass-metallicity and mass-sSFR relations in cluster and field}
In this section we present several aspects of galaxy evolution in different environments,  in the framework of the relations between mass and sSFR  and mass and gas phase metallicity. We  subdivide the samples to study the environmental  effects of the cluster and field populations. In order to be able to perform a meaningful investigation, and considering our relatively small number of targets, we divide the galaxies into three  mass-bins, like in \cite{maier16}:
\begin{enumerate}
  \item The low mass bin  $8.3<log(M/{\rm M}_{\odot})<9.2$ (50 R2248 cluster galaxies, 22 field galaxies). 
  \item  The intermediate mass bin  $9.2<log(M/{\rm M}_{\odot})<10.2$  (58 R2248 cluster galaxies, 66 field galaxies). This bin is well populated by both field and cluster galaxies, making it the mass complete bin. The comparison between the different populations in terms of their sSFRs and (O/H)s are based on galaxies from this intermediate mass bin.
  \item The high mass bin $10.2<log(M/{\rm M}_{\odot})<11.1$ (13 R2248 cluster galaxies, 40 field galaxies).
\end{enumerate}

\subsection{sSFR-Mass relation at $z\sim0.35$}
The SFR mass relation was investigated as first by \cite{noeske07} for a sample of galaxies in the All-Wavelength Extended Groth Strip International Survey field with redshifts between $0.2<z<1.1$. The results of this study have demonstrated that galaxies with reliable signs of SF form an apparent sequence with a limited range of SFR values at a given stellar mass and redshift (a scatter of $\sigma<0.35$ dex) and with $log(SFR) \propto M*$. This correlation between SF and stellar mass was defined as the "main sequence of star forming galaxies" (MS).\\
Fig. \ref{SFRM} shows the sSFR-M relation for the R2248 cluster members (filled circles) and for the comparison sample of CLASH and zCOSMOS field galaxies (open circles) with available measurements of the $H\alpha$ emission line. Both \textit{f=1} and  \textit{f=2} cluster galaxies are considered. The black cross in the upper left corner represents the mean error of the sSFR and M estimation. The red solid line  represents  the main sequence of star forming galaxies for a redshift of $z\sim0.4$, as derived by \cite{peng10} through the following equation: $log(sSFR)=-0.24\cdot log(M)+1.86$. The  red, dashed lines encompass the 0.3 dex dispersion of the MS. We have assumed a dependency of sSFR on mass as it was observed
for SDSS galaxies in the local universe by \cite{renzpeng15} : $ sSFR \propto m_{star}^{\beta} \: with \: \beta \sim  -0.24 $. \\
The red filled and blue open circles show the median values of the sSFRs of cluster and field galaxies, respectively.  The vertical, dashed lines show the 3 different mass bins, for which the median value of the sSFR was computed.  The mass complete sample is represented by  the intermediate bin ($9.2<log(M/{\rm M}_{\odot})<10.2$). In this mass bin, the median sSFR values of both cluster and field galaxies are comparable, given the errors.\\
Fig. \ref{DsSFR} displays the offset of the sSFRs of R2248 cluster members (filled circles) and the comparison sample of CLASH and zCOSMOS field galaxies (open circles)  to the main sequence of star forming galaxies at a redshift of $z\sim 0.4$,  as derived by \cite{peng10}.  The red filed and blue open symbols show the median values of   $\Delta sSFR$ for cluster and field galaxies respectively. \\
The median values for the sSFRs of both field and cluster galaxies and the scatter of the data points are quite consistent with the MS relation at  $z \sim 0.4$ and  with the scatter predicted by \cite{peng10}. By taking into account the mean error of the SF estimation of $\sim 0.3$ dex, the wast majority of the field and cluster galaxies can be classified as MS SF-galaxies. The error of the SFR was estimated through error propagation, by taking into account the flux measurement errors of the $H\alpha$ and  $H\beta$ ELs, together with the resulting error of the dust extinction from the Balmer decrement. The distribution  of both populations  in the mass-sSFR plane is comparable, especially in the  intermediate stellar mass bin $9.2<log(M/{\rm M}_{\odot})<10.2$, where the 2 samples have a uniform coverage. Since only SF galaxies are considered within this work, no substantial differences between the sSFRs of field and cluster galaxies can be observed. Therefore, the mass-sSFR relation of SF galaxies from this sample seems to be rather independent of whether the galaxies populate the field or cluster. These results are consistent with the findings of \cite{maier16} and \cite{maier19}, who observed a similar distribution for members of  CLASH and LoCuSS clusters and for field galaxies at $z<0.5$ in the mass-sSFR plane. 
\subsection{MZR at $z\sim0.35$}
The MZR determined from EL diagnostics was introduced as first by  \citet{trem04}, who found a tight correlation  of 0.1 dex between stellar mass and the gas-phase oxygen abundance, which extends over 3 orders of magnitude in stellar mass and a factor of 10 in oxygen abundance. \\
Fig. \ref{MZR} shows the MZR for R2248 cluster members (filled circles) and the comparison sample of CLASH  and zCOSMOS field galaxies (open circles) using the O3N2 metallicity calibration of  \cite{kewley13a}. All cluster members and field galaxies with available measurements of the $[OIII]$, $H\beta$, $[NII]$ and $H\alpha $ lines are considered.  The magenta diamonds and curves represent the local SDSS relation. We selected from the SDSS spectroscopic catalogs a sample of emission line galaxies with 0.04<z<0.08  as described in \cite{maier16}.  After excluding AGNs we recomputed the SDSS MZR using the O3N2 relation. The dotted magenta curves represent the   $1\sigma$ scatter of this relation. The MZR for the SDSS sample was extrapolated down to lower masses of $log(M/{\rm M}_{\odot})=8.3$ by assuming that the slope remains constant for $log(M/{\rm M}_{\odot})<9$. The black cross in the upper left corner represents the mean error of the (O/H) and M derivation.  The red filled and blue open circles give the median (O/H) values of cluster and field galaxies, respectively. The vertical dashed lines show the 3 different mass bins, for which the median (O/H) values were computed, with the mass complete sample represented by  the intermediate bin ($9.2<log(M/{\rm M}_{\odot})<10.2$).  In this mass bin, the median (O/H) of the cluster galaxies is by 0.065 dex higher than the median (O/H) of the field galaxies, yielding a 1.8 $\sigma$ significance. \\
Fig. \ref{DMZR} displays the offset of the gas phase metallicities of the R2248 cluster members (filled circles) and the comparison sample of CLASH  and zCOSMOS field galaxies (open circles) to the local SDSS MZR, using the O3N2 metallicity calibration of  \cite{kewley13a}. The red filled and blue open circles stand for the median value of the $\Delta (O/H)$ of cluster and field galaxies, respectively.\\
It is clear from these diagrams that both the cluster and field population at a redshift of z$\sim$0.35 have chemical abundances that differ from the abundances of the local  galaxies.  
However, the median value for the (O/H) of R2248 cluster  and field galaxies from  the intermediate and high mass bin   is consistent with the $1\sigma$ scatter of the local MZR.  The highest offsets to the local MZR occurs for the lowest mass cluster and field galaxies (8.3<$log(M/{\rm M}_{\odot})<9.2$), which generally show higher metallicities than the sample of galaxies from the local universe. The median value of the (O/H)s of both cluster and field galaxies at the low mass end is by $\sim 0.25$ dex higher than that of the SDSS sample.   At the highest masses (10.2<$log(M/{\rm M}_{\odot})<11.1$), however, this trend is  reversed, with most galaxies showing lower oxygen abundance values than the local SDSS sample, by $\sim 0.05$ dex.\\ When considering the mass complete bin $9.2<log(M/{\rm M}_{\odot})<10.2$, cluster galaxies show more enhanced metallicities than their field counterparts, with a 1.8 $\sigma$ significance.  This observed metallicity effect is investigated  more  thoroughly in the section that follows.

\section{Environmental effects and tentative evidence for starvation}
In this section we argue whether the ram pressure stripping of the hot halo gas is a plausible mechanism to account for the enhanced metallicities of the cluster members  from the mass complete bin, 
as compared to their field counterparts.

\subsection{MZR for cluster members with different accretion histories} 
In order to further explore the environmental effect on the gas regulation within galaxies, we compare  the median (O/H) values of the CLASH and zCOSMOS field  galaxies  to the median (O/H)s of  cluster members with different accretion histories. At cluster-centric distances lower than  $R_{500} \sim \frac{2}{3} R_{200}$, due to the high density of the ICM, cluster specific hydrodynamical processes such as ram pressure stripping should be more effective in removing the hot, diffuse halo gas of galaxies, especially if the galaxies have lower masses. Therefore,  it is expected that star forming galaxies, which are found at distances $R<R_{500}$, should have more enhanced metallicities than their counterparts residing at larger cluster centric distances and in the field, in accordance to the findings of for e.g. \cite{maier16} and  \cite{maier19}.  For this purpose, R2248 members located at $R<R_{500}$ were classified as "accreted" cluster galaxies and members residing at  $R>R_{500}$ were classified as the " infalling" population. \\
Fig. \ref{PSOH} displays on the left-hand side the cluster centric radius vs. line-of-sight velocity for the R2248 cluster members. The red symbols located at a cluster centric radius $R<R_{500}$ are considered to be "accreted" and possibly  virialised cluster members, which are in dynamical equilibrium with the clusters' gravitational potential. These systems were most probably accreted into the cluster at earlier epochs, and the passive, quenched population probably  passed through the apocentre of its first orbit.  According to \cite{haines15}, they identified this population of "accreted galaxies" as  galaxies that either formed locally or that  were accreted as the cluster’s core was being assembled.  The blue symbols located at a cluster centric radius $R>R_{500}$   are classified as "infalling" galaxies, which have been just recently accreted into the cluster. The ones showing high  line-of-sight velocities are  probably still on their first passage, and are not yet in dynamical equilibrium with the clusters' gravitational potential.  Some of these systems have probably just  recently passed within $R_{200}$ for the first time, but have not yet  reached the  pericentre. Therefore, the population of "infalling" galaxies shows overall higher line-of-sight velocities, as these systems are  accelerated when they travel deep into the gravitational potential well of the cluster core. The dashed black line shows $R_{500}$. The smaller, fainter points represent the entire sample of CLASH VIMOS R2248 galaxies, which were observed with a low resolution grism and for which no flux measurements are available. The dark grey-shaded area shows the $1\sigma$ contour of the caustic profile, while the light-grey  area shows the $2\sigma $ contour.\\ The panel from the right-hand side compares the MZR of the field galaxies, represented by the open black symbols, to the MZR of RXJ2248 cluster members with different accretion histories: the red symbols represent the cluster galaxies, which are classified as accreted and possibly virialised  members ($R<R_{500}$), while the blue symbols represent the "infalling" cluster members ($R>R_{500}$). To enable a meaningful comparison, only galaxies from the mass complete bin  ($9.2<log(M/{\rm M}_{\odot})<10.2$) are considered. The median values for the gas phase metallicities of the 3 populations are shown as the large symbols, colour coded accordingly. The magenta diamonds and curves show the local SDSS relation, as derived using the O3N2 method.\\  This investigation enables us to conclude the following: the median (O/H) value of the "accreted" cluster galaxies is by $0.04$ dex higher than that of the "infalling" cluster galaxies, but due to the low number of objects for the statistics, this difference yields only a $0.8\sigma$ significance. On the other hand,  the median (O/H) value of the "accreted" cluster galaxies is by $0.085$ dex higher than that of the field population, yielding a $1.9 \sigma$ significance. The distribution of the metallicities of $R < R_{500}$ galaxies  is clearly shifted to higher values compared to their field counterparts.\\
\cite{maier16} and \cite{maier19} observed a similar trend for other CLASH and LocuSS clusters, with galaxies at smaller cluster-centric radii showing more enhanced (O/H)s than galaxies located at larger cluster centric radii and in the field, and they concluded that infalling/recently accreted galaxies enhance their metallicities while travelling towards the clusters’ core. There are two possible scenarios to account for this: the systems either receive a metal enriched gas inflow into their halo (e.g. \cite{pengmaio14}, \cite{gupta18}), or the gas inflow rate into the galaxy is suppressed (e.g. \cite{pasq12}). Maier et al., however, showed that the observed metallicity trends can be explained by strangulation. 

\subsection{The fundamental metallicity relation Z(M,SFR) for the $z\sim 0.35$ galaxies}
As argued in the previous sections, it is clear that there exists a tight correlation between the SFR and  the stellar mass of a galaxy (sSFR-M relation), as well as a correlation between the gas-phase metallicity of a system  and its stellar mass (MZR). Observations have also shown that  oxygen abundances  anti-correlate with SFRs, especially at low stellar masses, such that at a given mass, galaxies with high SFR will show lower (O/H). This led to the conclusion that the chemical abundance of a galaxy is dependent on both the  stellar mass and the SFR, giving rise to the so-called "Fundamental Metallicity Relation" Z(M, SFR).\\
 For example \cite{mannu10}  assumed that an inflow of gas is the responsible driver for the increase of the SFR and the dilution of the metallicity in order to explain the star formation as a second parameter in the MZR. \cite{lilly13} proposed a different explanation to the dependency of Z on M and SFR, by introducing a simple model of galaxy evolution in which the SFR is regulated by the mass of gas present in a galaxy.\\
In this paper, we investigate whether at higher redshifts and in a dense cluster environment, the Z(M,SFR) is similar to the one found in the local universe. For this purpose, we calculate the expected (O/H) values from the simple gas regulated model of \cite{lilly13} for each galaxy individually, with their respective stellar mass  and SFR. This is done by means of equation (40) from  \cite{lilly13}, which can be rewritten as follows:
$$Z_{eq}=Z_{0}+y(1-R)SFR/\Phi $$
where $Z_{eq}$  is the equilibrium value for the metallicity, $Z_{0}$ the metallicity of the infalling gas, y the yield, R  the fraction mass returned to ISM, SFR the star formation rate and $\Phi $ the gas inflow rate. \\
E.g. \cite{mm19} have demonstrated, that both the shape and  the overall normalisation of the gas-phase MZR depend on the metallicity calibration used.  
Studies of blue (\cite{bresloin16}) and red supergiants (\cite{davies17}) have shown that the EL metallicity calibration, yielding absolute metallicities, is the O3N2 calibration, the one used throughout this work (cf. Fig. 15 from \cite{mm19}). Because of this, our measured  metallicities should be well suited to be compared, in an absolute
way, with the metallicities derived using the \cite{lilly13} bathtub model.
 To enable a meaningful comparison between theoretical models and observations, we compare our derived (O/H)s through the O3N2 method for galaxies from the mass complete bin to the predictions of \cite{lilly13}.  \\Fig. \ref{FMR} displays  the difference between the measured (O/H)s for  R2248 cluster galaxies (represented by the filled circles) and the comparison sample of field galaxies (represented by open circles), using the O3N2 metallicity calibration, and the expected (O/H)s from the FMR formulations of \cite{lilly13}. We use a model for primordial gas inflow: the metallicity of the infalling gas $Z_{0}$ relative to the yield y is taken to be $Z_{0}/y = 0$.
The colour bar displays the SFRs of the investigated galaxies in units $[{\rm M}_{\odot}/yr]$. The median values for the difference in (O/H) between measurements and predictions for both cluster and field galaxies are plotted as a reference and are represented by the big, dark and big, open circle respectively. The errors of the median are smaller than the symbols, and thus, can not be seen in the plot. Only galaxies from the mass complete sample ($9.2<log(M/{\rm M}_{\odot})<10.2$) are considered.  In this mass bin, the (O/H)s of the cluster galaxies deviate more strongly from the FMR model predictions than the ones for the field galaxies.\\  
The main results of the comparison between model and observations can be summarised as follows:  at intermediate masses, field galaxies seem to be in good agreement to the model predictions, showing a median deviation of $-0.01$  dex (big open circle in Fig. \ref{FMR}) from the expected (O/H) value. Cluster galaxies, on the other hand,  deviate more strongly from the theoretical predictions,  showing on average $ + 0.12$  dex (big dark circle  in Fig. \ref{FMR}) higher gas-phase metallicities than predicted.\\
Our results are consistent to the findings of \cite{maier16} and \cite{maier19} who studied the environmental effect on gas regulation within cluster galaxies and found that lower-mass cluster galaxies have higher metallicities than predicted by the models which assume an inflow of pristine gas, and that these systems also deviate more strongly from the FMR predictions than systems residing in lower density environments, indicating that a strangulation scenario, in which the gas inflow is cut off, can explain the enhanced (O/H)s. \\ 
These findings are also in accordance to the predictions of \cite{pengmaio14}, who studied the dependency of the MZR of SDSS galaxies on the environment, with regards to both the over-density and central/satellite dichotomy.  They  find that, for a given stellar mass,  there is a fairly strong dependency of the gas phase metallicity on the density of the ICM, with high metallicity galaxies (satellites) residing in denser regions than their low metallicity analogues.  The authors proposed  that the inflow of gas into the halo of a galaxy residing in a dense cluster environment should get progressively more metal-enriched (meaning that $Z_{0}$ becomes higher), producing thus enhanced metallicities for low mass cluster galaxies, as compared to their field counterparts. They also test their results by applying the gas-regulator model of \cite{lilly13} and find a good accordance with the models with metal rich gas inflow (also in agreement to the findings of e.g. \cite{gupta18} and  \cite{schaf19}). However,  no physical mechanism was proposed by \cite{pengmaio14} to explain how a metal enriched gas inflow  can still occur for galaxies residing in dense cluster environments. An alternative mechanism was proposed, which can lead to the same effects of enhanced metallicities in lower mass cluster galaxies, namely strangulation. This scenario is however not further elaborated,  due to the fact that they find that the sSFRs of cluster and field galaxies are similar. The same trend is observed for our investigated sample of galaxies too. \\  
Strangulation/starvation is a mechanism that causes the removal of the diffuse hot gas reservoir confined in the galaxy halo, while the gas-disc is left unperturbed.  This means that star formation can actually continue until the internal gas-reservoir is used up. Therefore, after the removal of the hot halo gas, there will be a time delay until the system will undergo star formation quenching. Because of this, even if the SFRs of cluster and field galaxies are comparable, that does not mean than strangulation should be excluded as a mechanism which explains the observed enhanced metallicities of intermediate mass cluster members, as compared to their field counterparts\\

\subsection{Discussion: environmental effects }
In a recent study, \cite{haines13} investigated a large sample of LoCuSS cluster galaxies of intermediate redshifts in terms of the evolution of their star formation activity, and found that massive, actively star forming galaxies, when accreted into high mass clusters, are slowly quenched. Infalling spiral galaxies will lose their gas reservoirs due to (mild) ram pressure stripping  as they interact with the hot ICM  and this will result in a gradual decrease of their SFRs. The authors found that the SFRs of cluster galaxies decline exponentially on time-scales between 0.7–2.0 Gyr, and concluded that  mechanisms such as ram pressure stripping and strangulation are responsible for quenching star formation on such long time scales.\\
This is in accordance with another recent study by \cite{jaffe15}, who investigated the effects of ram-pressure on  stripping the HI gas from intermediate z cluster galaxies from the  BUDHIES survey. They found  that ram-pressure plays an important role in removing the gaseous reservoir of galaxies, and that this can happen during the first infall into the cluster.  However, these galaxies are still able to continue to form stars from the remaining gas
disc. They also conclude that the gas stripping will occur once the galaxies have approached the dense ICM core and/or gained enough velocity to cross the “stripping” area in phase-space, defined as the small clustercentric distances and/or high line-of-sight velocities. After this stripping process, galaxies are expected to "oscillate" in phase-space,  until they accumulate towards the clusters' core due to dynamical friction. During this time, it is expected that the galaxies exhaust their  gas reservoirs, which will, in turn, cause the quenching of  star formation.\\
\cite{gg72} demonstrated that gas can be removed from galaxies, if the ram pressure exceeds the restoring force per unit area (i.e. gravitational restoring force) exerted by the galaxy. Based on this, \cite{bahe13} studied the environmental effects which lead to star formation quenching  in cluster galaxies employing  a suite of high-resolution cosmological hydrodynamic simulations. The authors of this paper have analysed galaxies from simulated groups and clusters with a wide range of stellar masses, in order to compute the density of the ICM and the ram pressure in clusters, and this allowed them to  investigate how effective ram pressure can strip away the gas reservoirs of galaxies.
They find  a systematic reduction of both the hot and cold gas component  and
a decline in the star forming fraction of galaxies with decreasing clustercentric distance.
For massive clusters  with $M \sim 10^{15} {\rm M}_{\odot}$  (similar  to the mass of the R2248 cluster),  Bah\'e et al. estimated the typical values for the ram pressure and those for the restoring pressure of the cold gas and warm gas. 
When comparing  the restoring pressure to the ram pressure near $R_{200}$, the latter one is too low to strip the cold gas in the most massive galaxies, but is just sufficiently high to strip the hot halo gas  from lower mass systems. As the hot halo gas is  less tightly gravitationally  bound (by c.a. two orders of magnitudes)  than the cold gas component, ram pressure can strip it off more efficiently, meaning that even the most massive galaxies can be affected  out to $\sim 2
- 3 R_{200}$ in clusters and most low mass  systems are subject to sufficient  ram pressure stripping of the hot halo gas  even at large clustercentric distances of $ 5 \cdot R_{200}$. \\
The removal of the hot gas component at $R> R_{200}$ due to strangulation will lead to a delayed decrease in the SFR of the system, as the cold gas disc remains unperturbed, and thus stars can  continuously form until the gas reservoir is consumed. Strangulation is a slow mechanism which at first, will not affect the SFR of the galaxy. However, as galaxies continue to move through the hot ICM towards the  dense cluster centre, the ram pressure will increase to values  which are comparable to the restoring pressure of the cold gas, such that the cold-gas component can also be stripped away. The removal of the cold gas disc of a galaxy will lead to a rapid phase of complete star formation quenching (see slow-then-rapid quenching scenario of \cite{maier19}).\\
Once a galaxy's gas supply is cut off due to strangulation, the system can be described through a "closed-box model". E.g. \cite{maier06} explored  which region of the parameter space could reproduce the constraints imposed by the metallicity-luminosity relation of both local and high z galaxies, based on a large grid of P{\'e}gase2 models. They analysed different models in which the gas supplies of the model galaxies are cut off, and these systems continue to form stars in a closed-box like environment. The tracks of these closed box models have demonstrated that after the gas inflow is suppressed and eventually stopped, galaxies will enhance their gas phase metallicities by a factor of $\sim 0.2$ dex on time scales of $\sim 1$ Gyr. According to these findings, cluster galaxies, which experience strangulation, can enhance their metallicities while travelling towards the central parts of the cluster. \\
The observational results of this work reveal enhanced metallicities for intermediate mass cluster galaxies, as compared to the population of field galaxies at similar redshifts.
As the investigated R2248 cluster galaxies extend out to a maximum distance from the cluster centre of $\sim 1.5 R_{200}$,  ram pressure stripping has probably already affected the hot halo gas of  these systems. Because of this, no significant  difference (a difference of 0.04 dex yielding only $0.8\sigma$ significance)  can be observed in  the (O/H)s of "infalling" and "accreted" cluster members, as both of these populations reside, in projection, close to the cluster core. However, as seen in Fig. \ref{MZR}, the combined population of "accreted" and "infalling" cluster galaxies shows more enhanced metallicities than the field, with a $\sim 1.8 \sigma$ significance. 
This indicates that,  when accreted,  cluster galaxies (especially lower mass ones) are more prone to be affected by strangulation, and therefore, they will increase their metallicities while moving through the hot ICM towards the central regions of the cluster. The observed metallicity effect is indeed weak, but this is also related to the fact that a low number of galaxies was used for the statistics. To conclude, ram pressure stripping of the hot halo gas can be considered as a plausible mechanism to account for  the observed metallicity effects in cluster galaxies of intermediate redshifts. \\

\section{Summary and Conclusion}
\begin{figure*}
\begin{subfigure}
    \centering
    \includegraphics[width=0.5\textwidth,angle=0,clip=true]{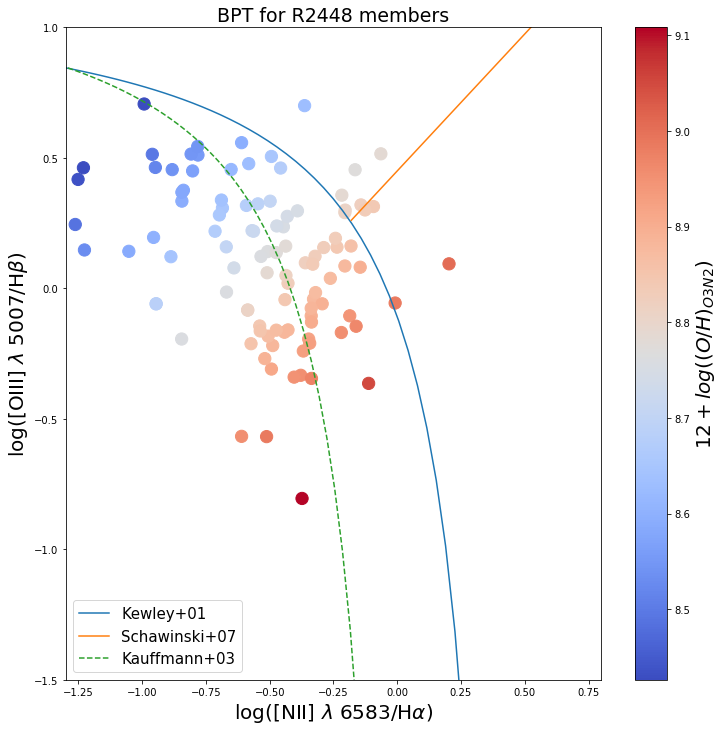}
    \centering
\end{subfigure} 
\begin{subfigure}
    \centering
    \includegraphics[width=0.5\textwidth,angle=0,clip=true]{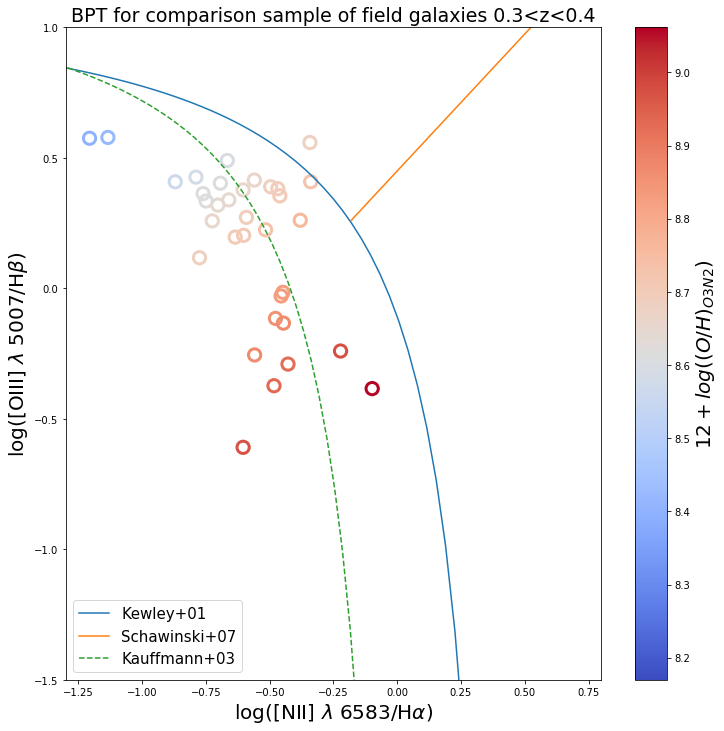}
    \centering
\end{subfigure} 
 \caption{BPT  (\cite{bald81}) diagram for the R2248 cluster members (\textit{left}) and comparison sample of CLASH  field galaxies with $0.3<z<0.4$ (\textit{right}). This diagnostic diagram is used to distinguish whether the dominant source of ionisation comes from the stellar component or is rather associated with an AGN. The solid, blue curve represents the theoretical curve of \cite{kewley01} and the red, dashed one represents the empirical curve of \cite{kaufm03}, which separate star- forming galaxies (below/left of the curves) from AGNs (above/right of the curves). The green solid curve was derived by \cite{sw17}  as the separation between Seyfert II galaxies (above and to the left) and LINERs (below and to the right).  The colour bar shows the oxygen abundances of the investigated galaxies, as computed through the  O3N2 method of \cite{kewley13a}. All CLASH galaxies classified as AGNs/LINERs were excluded from the sample.}  
\label{BPTfc}
\end{figure*}

\begin{figure*}
   \centering
    \includegraphics[width=0.7\textwidth,angle=0,clip=true]{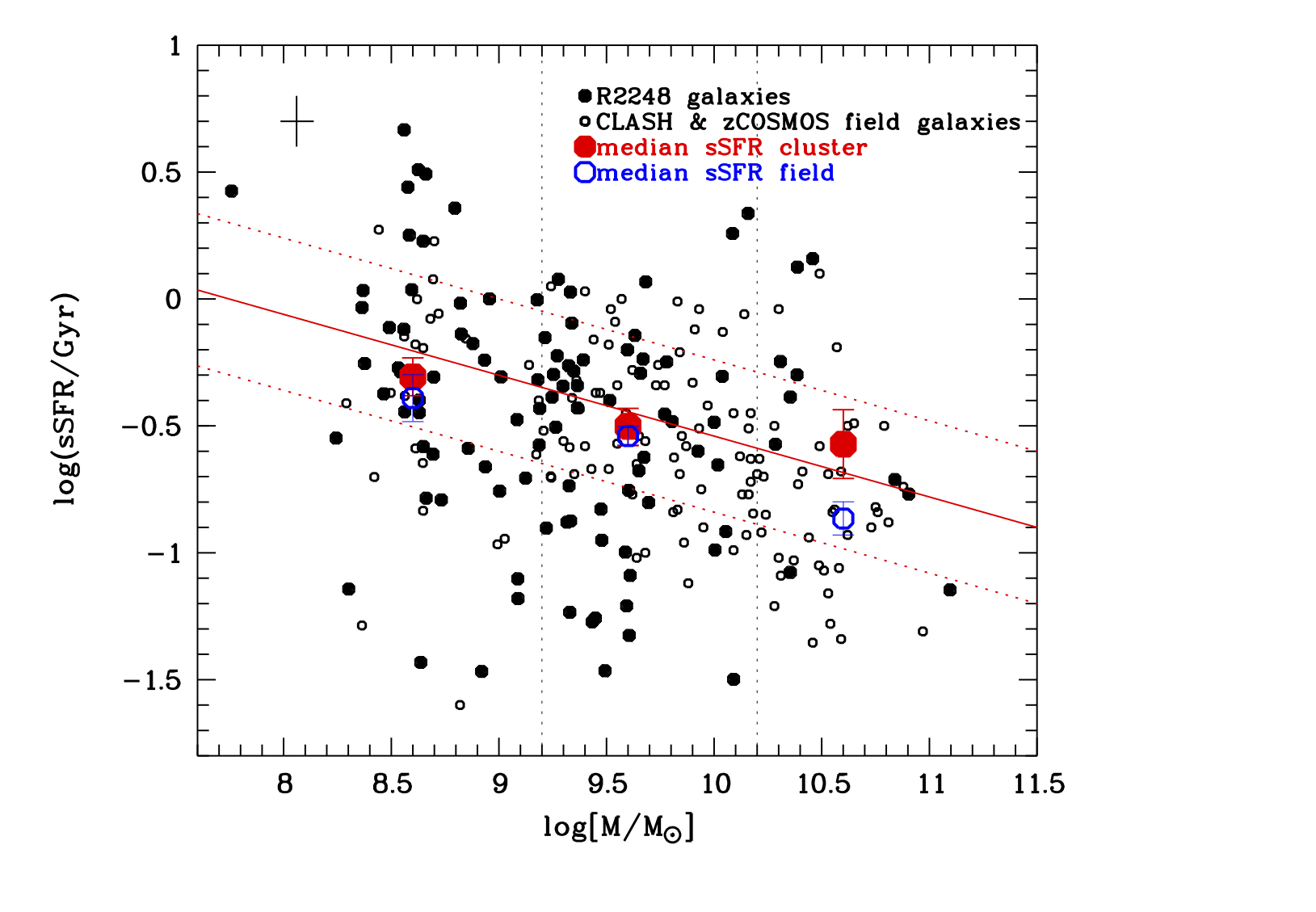}
    \centering
  \caption{sSFR-M relation for the R2248 cluster members (filled circles) and for the comparison sample of CLASH and zCOSMOS field galaxies (open circles) with available measurements of the $H\alpha$ emission line. Both \textit{f=1} and  \textit{f=2} galaxies are considered. The median values of the sSFRs of  cluster galaxies are plotted as the  red symbols, while the blue, open symbols represent the median values of the sSFR of the field galaxies. The median sSFR values were computed for the 3 different mass bins, which are  represented by the vertical, dashed lines. The red solid line stands for the the main sequence of star forming galaxies for a redshift of $z\sim0.4$, as derived by \cite{peng10}. The dashed red lines encompasses the 0.3 dex dispersion of the MS. The black cross in the upper left corner  shows the mean error of the sSFR and M estimation.  The mass complete sample is represented by  the intermediate bin ($9.2<log(M/{\rm M}_{\odot})<10.2$). In this mass bin, the median sSFR values of both cluster and field galaxies are comparable, given the errors.\\  }
 \label{SFRM}%

\end{figure*}

\begin{figure*}
    \centering
    \includegraphics[width=0.7\textwidth,angle=0,clip=true]{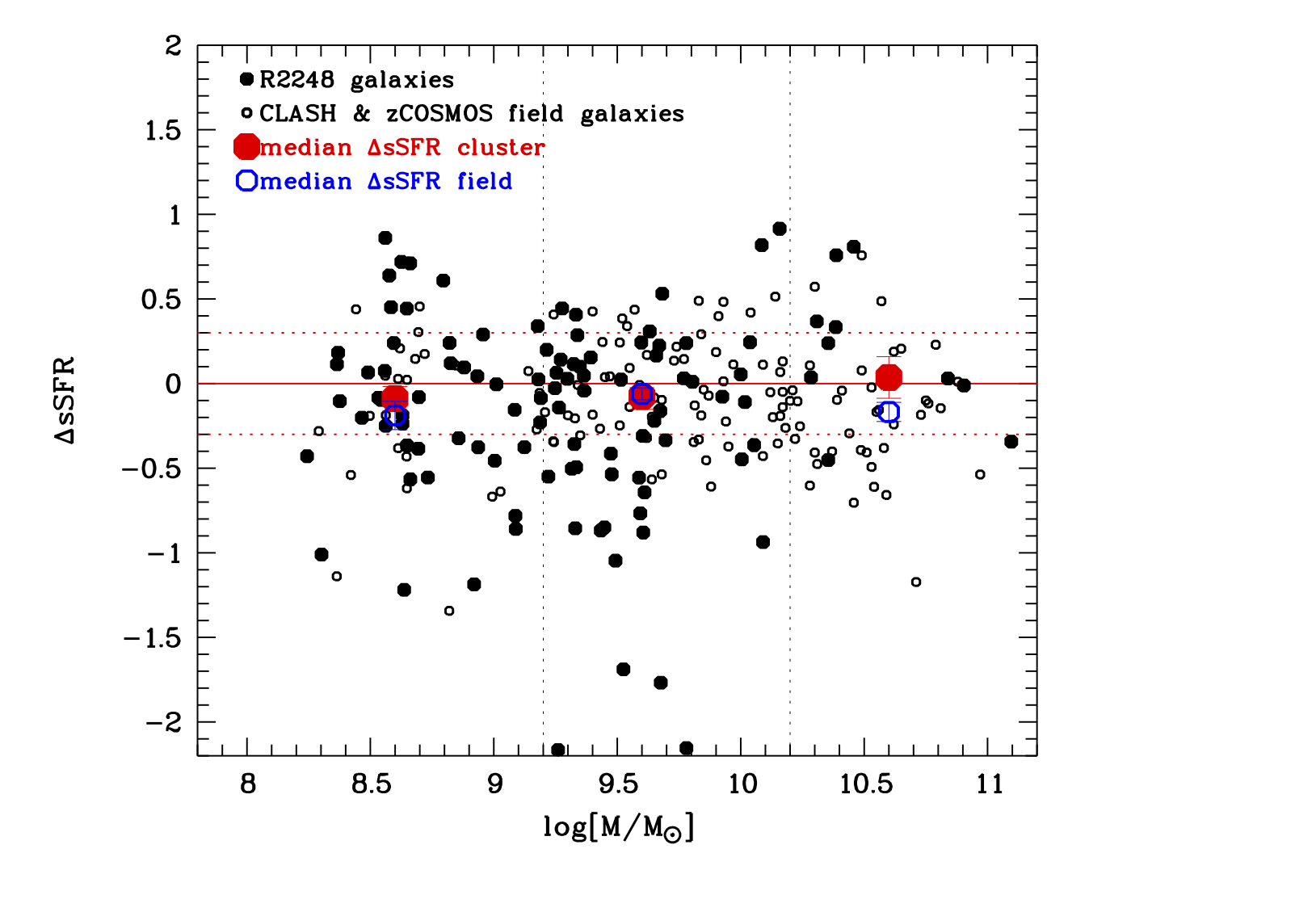}
    \centering
 \caption{Offset of the sSFRs of R2248 cluster members (filled circles) and the comparison sample of CLASH and zCOSMOS field galaxies (open circles)  to the main sequence of star forming galaxies at a redshift of $z\sim 0.4$,  as derived by \cite{peng10} .  The red  filed and blue open symbols indicate the median values of   $\Delta sSFR$ for cluster and field galaxies respectively, as computed for the 3 different mass bins.}
\label{DsSFR}
\end{figure*}

\begin{figure*}
   \centering
    \includegraphics[width=0.7\textwidth,angle=0,clip=true]{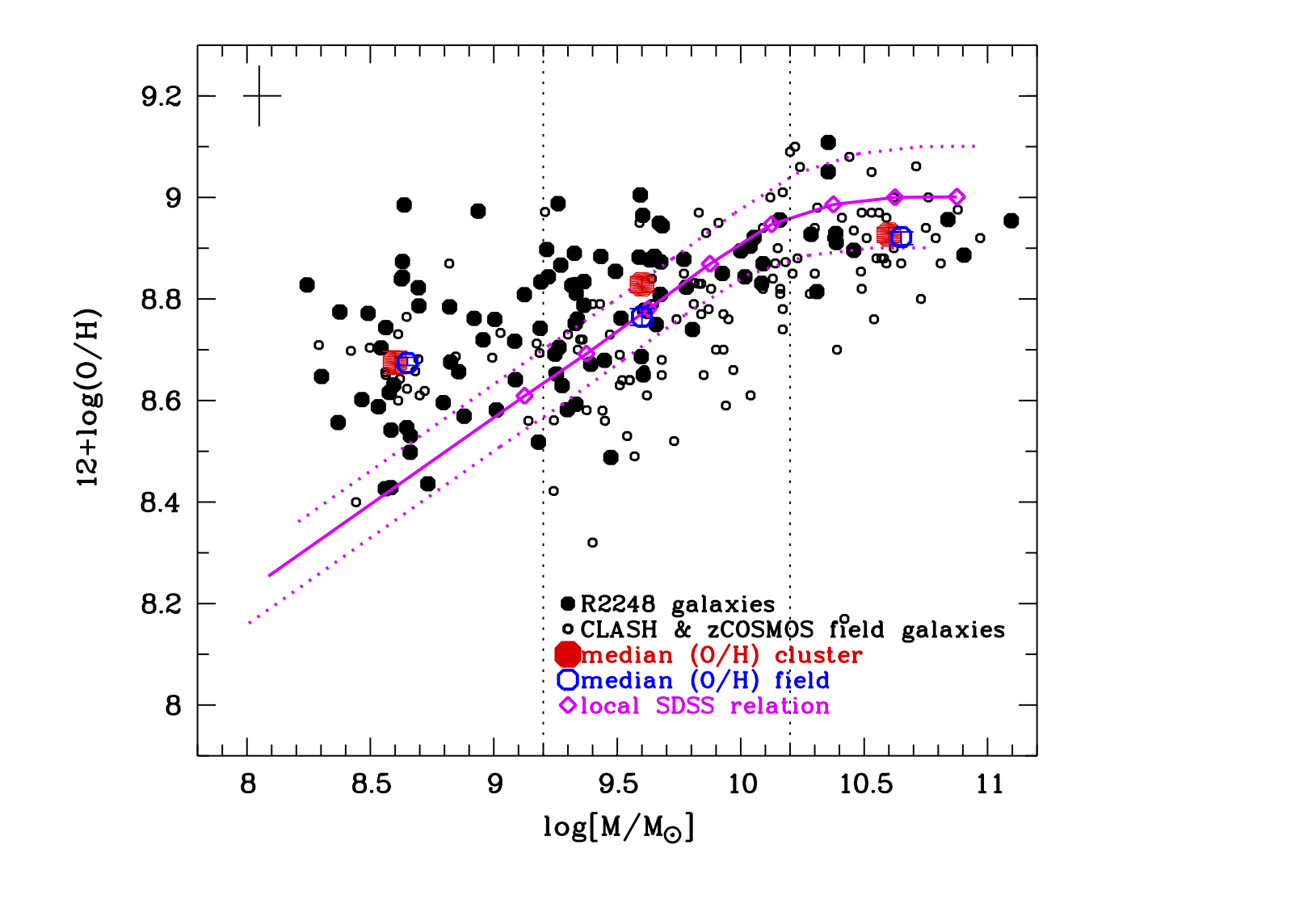}
    \centering
 \caption{Mass-metallicity relation for R2248 cluster members (filled circles) and the comparison sample of CLASH  and zCOSMOS field galaxies (open circles) using the O3N2 metallicity calibration of  \cite{kewley13a}. All cluster members and field galaxies with available measurements of the $[OIII]$, $H\beta$, $[NII]$ and $H\alpha $ lines are considered.   The magenta diamonds and curves represent the  local SDSS relation,  consistently derived using the O3N2 calibration, with its $1\sigma$ scatter. The red filled and blue open circles show the median (O/H) values of cluster and field galaxies, respectively. The black cross in the upper left corner represents the mean error of the (O/H) and M derivation. The vertical, dashed lines show the 3 different mass bins, for which the median (O/H) values were computed, with the mass complete sample represented by  the intermediate bin ($9.2<log(M/{\rm M}_{\odot})<10.2$). In this mass bin, the median (O/H) for the cluster galaxies is by 0.065 dex higher than the median (O/H) for the field galaxies, with a 1.8 $\sigma$ significance.}
\label{MZR}
\end{figure*}

\begin{figure*}
    \centering
    \includegraphics[width=0.7\textwidth,angle=0,clip=true]{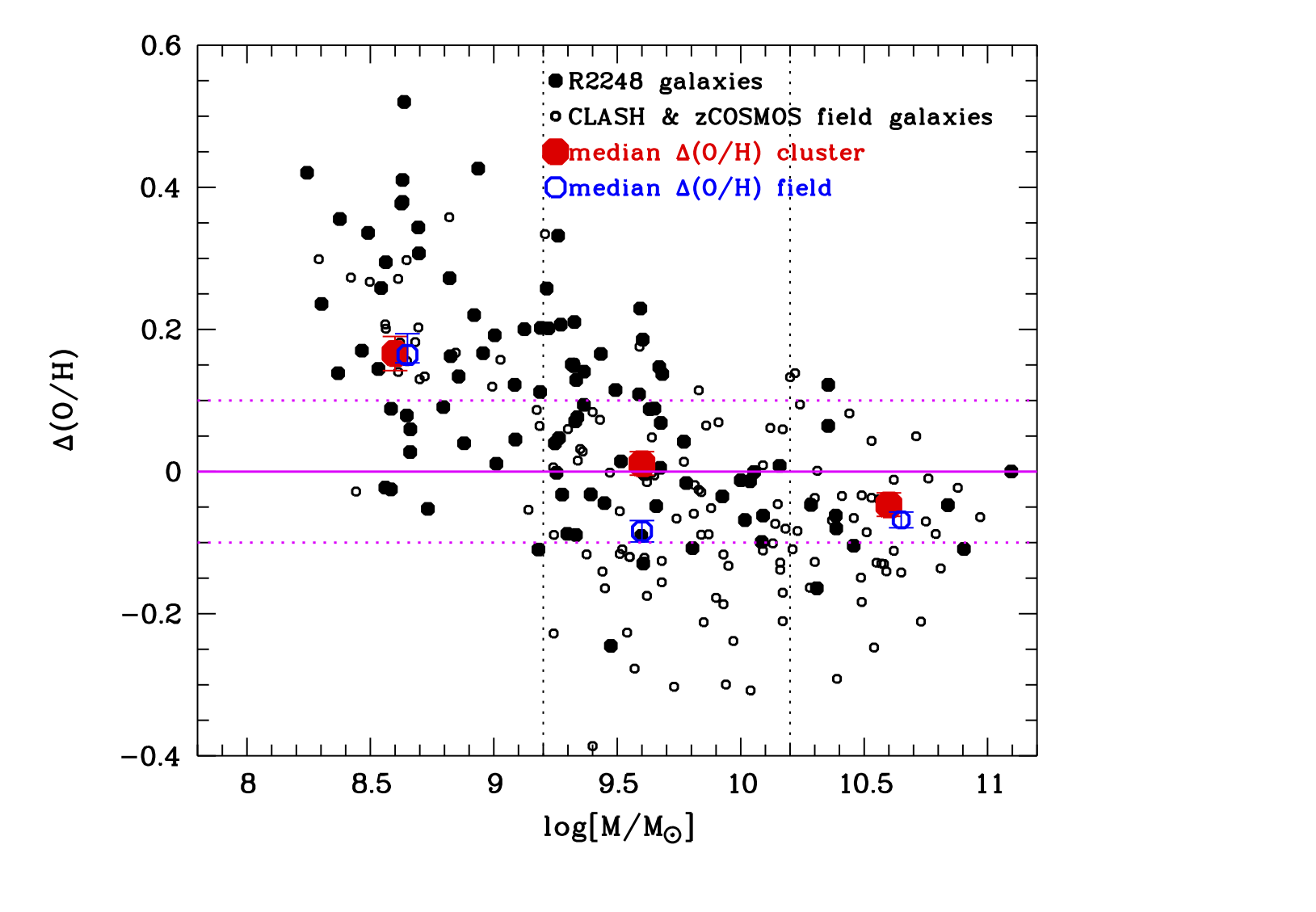}
    \centering
 \caption{Offset of the gas phase metallicities of the R2248 cluster members (filled circles) and the comparison sample of CLASH  and zCOSMOS field galaxies (open circles) to the local SDSS MZR, using the O3N2 metallicity calibration of  \cite{kewley13a}. The red filled and blue open circles show the median value of the $\Delta (O/H)$ of cluster and field galaxies respectively, as computed for the 3 different mass bins.}
\label{DMZR}
\end{figure*}

\begin{figure*}
\begin{subfigure}
    \centering
    \includegraphics[width=0.6\textwidth,angle=0,clip=true]{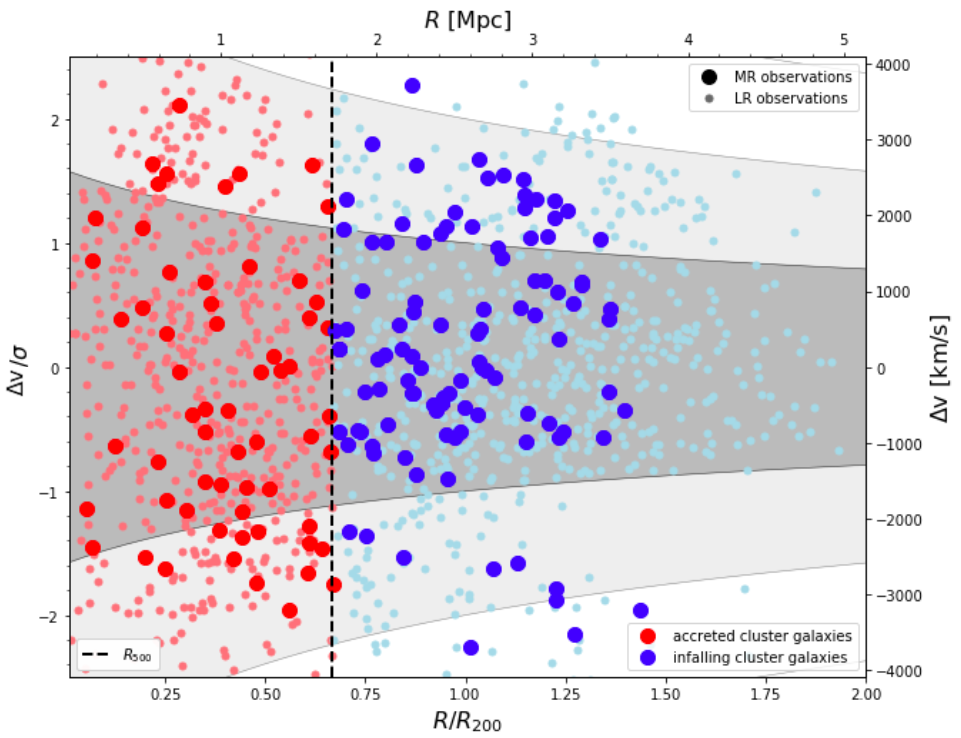}
        \centering
\end{subfigure} 
\begin{subfigure}
    \centering
    \includegraphics[width=0.55\textwidth,angle=0,clip=true]{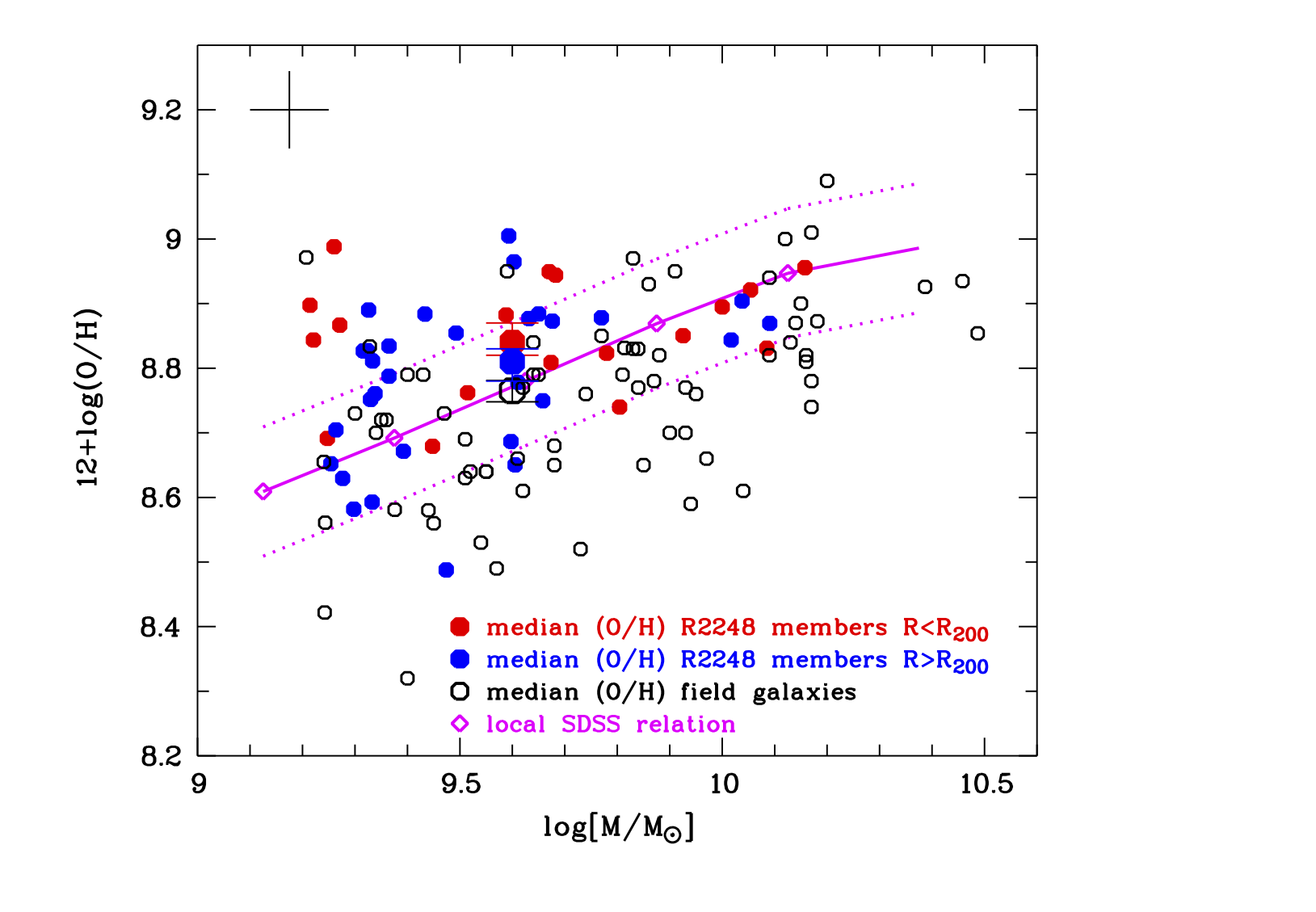}
    \centering
\end{subfigure} 
 \caption{ \textit{Left}: Cluster centric radius vs. line-of-sight velocity for the R2248 cluster members. The red symbols located at a cluster centric radius $R<R_{500}$ are considered to be "accreted" and possibly  virialised cluster members, which are in dynamical equilibrium with the clusters' gravitational potential. The blue symbols located at a cluster centric radius $R>R_{500}$   are classified as "infalling" galaxies, which have been just recently accreted into the cluster. The dashed black line stands for $R_{500}$. The large points represent the galaxies with middle resolution observations, while the smaller, fainter points represent the entire sample of CLASH VIMOS R2248 galaxies, which were observed with a low resolution grism and for which no flux measurements are available. The dark grey-shaded are shows the $1\sigma$ contour of the caustic profile, while the light-grey shaded area shows the $2\sigma $ contour. \\ \textit{Right}:  MZR comparison between field galaxies, represented by the black open symbols, and the R2248 cluster members with different accretion histories: the red and blue symbols represent the "accreted" and "infalling" cluster members, respectively. Only galaxies from the mass complete bin ($9.2<log(M/{\rm M}_{\odot})<10.2$) are considered. The median values for the gas phase metallicities of the 3 samples are shown as the large symbols, colour coded accordingly. The magenta curves and diamonds show the local SDSS relation, as derived using the O3N2 calibration.  The median (O/H) value of the "accreted" cluster galaxies is by $0.04$ dex higher than that of the "infalling" cluster galaxies, but due to the low number of  members for the statistics, this difference yields only a $0.8\sigma$ significance. On the other hand,  the median (O/H) value of the "accreted" cluster galaxies is by $0.085$ dex higher than that of the field population, yielding a $1.9 \sigma$ significance. }
\label{PSOH}
\end{figure*}

\begin{figure*}
    \centering
    \includegraphics[width=0.7\textwidth,angle=0,clip=true]{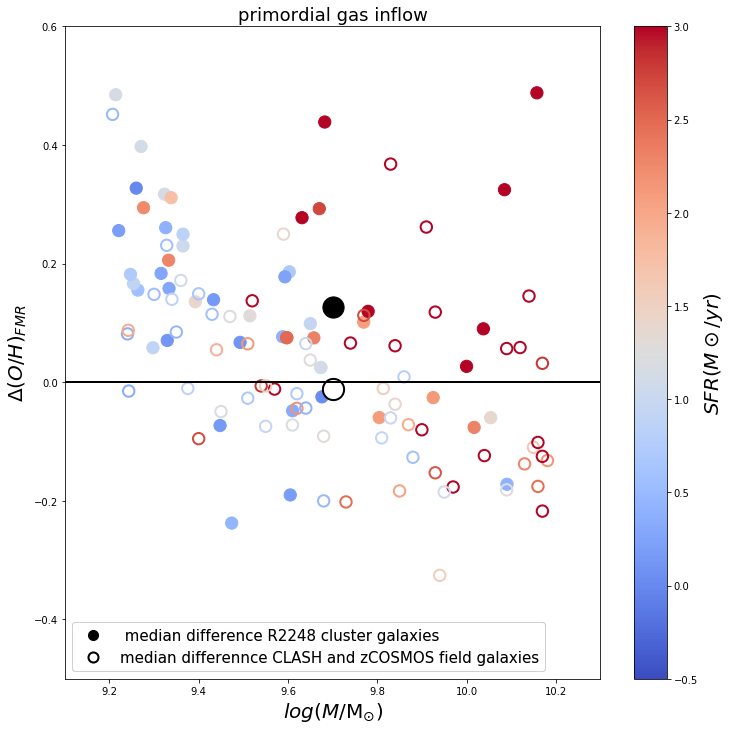}
    \centering
 \caption{Difference between the measured (O/H)s for  R2248 cluster galaxies (represented by the filled circles) and the comparison sample of field galaxies (represented  by the open circles), using the O3N2 metallicity calibration, and the expected (O/H)s from the formulations of \cite{lilly13} for a primordial infall metallicity $Z_{0}$ relative to the yield y, $Z_{0}/y = 0$.
  As a third dimension, the SFRs of the investigated galaxies are shown in units $[{\rm M}_{\odot}/yr]$. The median values for the difference in (O/H) between measurements and predictions for both  cluster and field galaxies are plotted as a reference and are represented by big dark and big open circles respectively. The errors of the median are smaller than the symbols, and thus, can not be seen in the plot.  Only the galaxies from the mass complete bin ($9.2<log(M/{\rm M}_{\odot})<10.2$) are considered for this investigation.  In this bin, the (O/H)s of the cluster galaxies (big filled circle) deviate more strongly from the FMR model predictions  (by $\sim 0.12$ dex) than the ones for the field galaxies (big open circle), which are in quite good agreement to the model predictions.
}
\label{FMR}
\end{figure*}

We study the impact that the environment has on the Z( M, SFR) relation of both CLASH R2248 cluster members at $z\sim0.35$ and of field galaxies with $0.3<z<0.413$ based on CLASH-VLT VIMOS spectroscopy.
The main results can be summarised as follows:
\begin{enumerate}
 \item  \textit{ SF galaxies and AGNs:} We have used the BPT diagnostic diagram to differentiate between the main sources of ionisation within galaxies. Both field and cluster galaxies are not dominated by AGNs, and they follow the SF-sequence in the BPT diagram (Fig. \ref{BPTfc}).
\item \textit{sSFR-M relation:} Both field and cluster galaxies can be classified as SF "Main Sequence" galaxies, with just a lower fraction of galaxies classified as starburst or passive systems. Field and cluster galaxies show a similar  sSFR-M and we therefore concluded that this relation seems to be independent of environment, when considering only SF galaxies. (Fig. \ref{SFRM}). 
\item \textit{MZR:} Both field and cluster galaxies follow the MZR of local SDSS galaxies, with an offset of low mass galaxies ($8.3<log(M/{\rm M}_{\odot})<9.2$) towards higher metallicities than the local MZR. Cluster galaxies with $9.2<log(M/{\rm M}_{\odot})<10$ show more enhanced metallicities than their field counterparts, by $\sim 0.065$ dex, yielding a $\sim 1.8 \sigma$ significance (Fig. \ref{MZR}).
\item \textit{Tentative evidence for strangulation: }
We find several indications for a strangulation scenario for
 cluster galaxies with $9.2<log(M/{\rm M}_{\odot}) < 10.2$
\begin{enumerate}
\item \textit{(O/H) comparison between field and cluster galaxies at different clustercentric radii:} In the mass complete bin, cluster galaxies with $R<R_{500}$
 show more enhanced metallicities than their counterparts residing at $R>R_{500}$, by 0.04 dex, but due to the low number of objects for the statistics, this difference yields a $0.8\sigma$ significance, and thus, their metallicities can be considered to be consistent. The fact that we observe no significant difference  between the metallicities of accreted and "infalling" cluster members can be explained by the fact that the sample of cluster galaxies resides, in projection, close to the cluster core (all within $1.5 \cdot R_{200}$), meaning that cluster specific processes, such as ram pressure stripping, have probably affected the hot halo gas of all these galaxies. On the other hand,  the median (O/H) value of the "accreted" cluster galaxies is by $0.085$ dex higher than that of the field population, yielding a $1.9 \sigma$ significance.
(Fig. \ref{PSOH}). 
\item \textit{FMR:}  In the mass complete bin, the measured (O/H)s of field galaxies seem to be in better agreement with the expected (O/H)s from the gas-regulated model of \cite{lilly13}, than the (O/H)s of cluster galaxies, which deviate more strongly from the model predictions, by 0.12 dex (Fig. \ref{FMR}).  This discrepancy between observed and predicted metallicities indicates that starvation can indeed be responsible for the  enhanced metallicities of cluster galaxies. 

\end{enumerate}

\end{enumerate}
To conclude, the observational results of this work favour the scenario in which  ram pressure stripping is responsible for the removal of the hot halo gas of R2248 cluster galaxies with  $9.2<log(M/{\rm M}_{\odot}) < 10.2$, thereby increasing their (O/H)s. We should also mention that the observed metallicity effects are indeed weak, but this is also due to the fact that we had a low number of objects with middle-resolution observations for the statistics. All cluster members, are located close to the cluster core and do not reach the outskirts of the caustic profile, and thus, a meaningful comparison between galaxies of different accretion histories could not be carried out.\\

\begin{acknowledgements}

We would like to express our deep gratitude to Amata Mercurio and  Piero Rosati for providing us with both the VIMOS redshift catalogue and  the WFI photometric catalogue for the R2248 cluster galaxies with middle resolution VIMOS observations. We also thank the CLASH-VLT team for the data reduction. Special thanks to José Manuel Perez Martinez for sharing his expertise regarding LePhare with us. We would also like to thank the anonymous referee for providing constructive comments and help in improving the manuscript.
\end{acknowledgements}



%
%




\newpage

\begin{table*}[b]
\caption{Catalogue for the RXJ2248 cluster galaxies: fluxes of the ELs $H\beta $, [OIII], $H\alpha$, [NII]  in units $10^{-17}  erg/s/cm^2/\AA$
 together with the errors of the flux measurements, the stellar mass as computed by LePhare, the SFR in  $log(sSFR/Gyr)$, as computed using the luminosity of the $H\alpha$ EL and the oxygen abundance derived through the O3N2 method. Galaxy IDs, coordinates and redshift will be available in a future publication  (Mercurio et al. in prep.). \emph{The complete version of this table (now only three objects) for the full sample of about 178 galaxies will be available after the paper is published in A$\&$A}.}

\begin{adjustwidth}{-0.5in}{-0.5in}
\begin{center}
\begin{adjustbox}{width=1\textwidth}
\small
\begin{tabular}{lcccccccl}
\hline
Galaxy  	&	$F(H\beta) $ 	&	$F([OIII])$ 	&	$F(H\alpha) $ 	&	$F([NII])$		&	 $log(M/{\rm M}_{\odot})$ &  $log(sSFR/Gyr)$ & $12+log(O/H)$    	\\ 						
\hline
 01	&	 1.13 $	\pm	$ 0.12 	&	 2.77 $	\pm	$ 0.18 	&	  4.03 $	\pm	$ 0.15  	&	 1.28 $	\pm	$  0.30 	&	  8.54 	&	 -0.29  	&	 8.70     	\\
 02	&	    * $	\pm	$    * 	&	 0.58 $	\pm	$ 0.10 	&	     * $	\pm	$    *  	&	 0.67 $	\pm	$  0.08 	&	  9.57 	&	     *  	&	    *	\\
 03	&	    * $	\pm	$    * 	&	 1.77 $	\pm	$ 0.25 	&	     * $	\pm	$    *  	&	    * $	\pm	$     * 	&	 10.61 	&	     *  	&	    *	\\
 \end{tabular}
\end{adjustbox}
\end{center}
\end{adjustwidth}
\label{tab}
\end{table*}

\end{document}